\documentclass[iop]{emulateapj}
\usepackage{natbib,rotating,enumitem,amsmath}

\newcommand{\msol}{\,\textrm{M}_\sun}                
\newcommand{\ATD}{ATLAS${}^{\rm3D}$}
\setlist[enumerate]{noitemsep}
\accepted{5 June 2018}
\begin{document}

\title{Resolving Quiescent Galaxies at $z \gtrsim 2$: II. Direct Measures of Rotational Support}
\shorttitle{Resolving Quiescent Galaxies at $z > 2$: II.}
\shortauthors{Newman, Belli, Ellis, and Patel}
\author{Andrew B. Newman$^1$, Sirio Belli$^2$, Richard S. Ellis${}^{3,4}$, and Shannon G. Patel$^1$}
\affil{$^1$ The Observatories of the Carnegie Institution for Science, Pasadena, CA, USA; \href{mailto:anewman@carnegiescience.edu}{anewman@carnegiescience.edu}}
\affil{$^2$ Max-Planck-Institut f\"ur Extraterrestrische Physik (MPE), Giessenbachstr.~1, D-85748 Garching, Germany}
\affil{$^3$  Department of Physics and Astronomy, University College London, Gower Street, London WC1E 6BT, UK}
\affil{$^4$ European Southern Observatory (ESO), Karl-Schwarzschild-Strasse 2, D-85748 Garching, Germany}

\begin{abstract}
Stellar kinematics provide insights into the masses and formation histories of galaxies. At high redshifts, spatially resolving the stellar kinematics of quiescent galaxies is challenging due to their  compact sizes. Using deep near-infrared spectroscopy, we have measured the resolved stellar kinematics of four quiescent galaxies at $z=1.95$-2.64, introduced in Paper~I, that are gravitationally lensed by galaxy clusters. Analyses of two of these have previously been reported individually by Newman et al.~and Toft et al., and for the latter we present new observations. All four galaxies show significant rotation and can be classified as ``fast rotators.'' In the three systems for which the lensing constraints permit a reconstruction of the source, we find that all are likely to be highly flattened (intrinsic ellipticities of $\approx0.75$-0.85) disk-dominated galaxies with rapid rotation speeds of $V_{\rm max}=290$-352~km~s${}^{-1}$ and predominantly rotational support, as indicated by the ratio (V/$\sigma)_{R_e}=1.7$-2.3. Compared to coeval star-forming galaxies of similar mass, the quiescent galaxies have smaller $V/\sigma$. Given their high masses $M_{\rm dyn}\gtrsim2\times10^{11} \msol$, we argue that these galaxies are likely to evolve into ``slow rotator'' elliptical galaxies whose specific angular momentum is reduced by a factor of 5-10. This provides strong evidence for merger-driven evolution of massive galaxies after quenching. Consistent with indirect evidence from earlier morphological studies, our small but unique sample suggests that the kinematic transformations that produced round, dispersion-supported elliptical galaxies were not generally coincident with quenching. Such galaxies probably emerged later via mergers that increased their masses and sizes while also eroding their rotational support.
\end{abstract}


\section{Introduction}

Modern theories of massive galaxy formation generally posit two broad evolutionary phases \citep[e.g.,][]{Oser10}. In the first phase at high redshifts, a highly dissipative event (e.g., a major merger or disk instability; \citealt{Zolotov15}) leads to the formation of a compact quiescent galaxy. Such galaxies do not resemble fully formed ellipticals, since they have sizes that are too small for their mass \citep{Trujillo06,vanDokkum08}. In the second phase, extended stellar ``wings'' gradually emerge around the compact core and the half-light radius increases \citep{vanDokkum10,Patel13}. This growth is unaccompanied by significant star formation and is thought to arise primarily from the accretion of low-mass satellites \citep{Naab09,Bezanson09}, although secondary effects including residual star formation and the expansion of stellar orbits may also contribute \citep{Hopkins10,Newman12,Wellons15}.

Although the compact sizes of massive quiescent galaxies at $z \gtrsim 2$ received much of the initial attention, more detailed information about their internal structure has been emerging over the last decade. Contemporaneous with the discovery of the compact sizes of this population, it was noticed in small samples of galaxies that some have disk-like morphologies and surface brightness profiles \citep{Stockton08,McGrath08,vanDokkum08,vanderWel11}. Larger samples confirmed that a substantial fraction of massive quiescent galaxies at $z \sim 2$ appear to be disk dominated, although quantitative estimates of this fraction varied considerably \citep{Bruce12,Bruce14,Buitrago13,McLure13}. Based on viewing angle arguments and the evolving axis ratio distribution, which shows that an increasing fraction of massive quiescent galaxies have flattened shapes toward higher redshifts, \citet{Chang13a,Chang13b} claimed that the majority are likely to be disk-dominated.

Altogether these studies have provided mounting photometric evidence that many quiescent galaxies---including the most massive examples that are predominantly round and dispersion supported today---were more disk-like at early epochs. This has important implications for the formation and evolution of massive galaxies: it would imply that most of the stars formed in a disk that survived whatever processes quenched star formation, and that quiescent galaxies not only grew in size after quenching but also underwent major changes in their shapes and distribution of stellar orbits.

Kinematic evidence is necessary to definitively identify disk-dominated structures in early quiescent galaxies. Using ground-based spectroscopy, it would be extremely difficult to directly measure rotation in distant quiescent galaxies because of their small angular sizes. Unresolved kinematics have provided indirect evidence of rotation. In a study of 24 quiescent galaxies at $z=1$-2.5 with measured velocity dispersions, \citet{Belli17} showed that within the subsample classified as disky based on their S\'{e}rsic indices, the ratio of dynamical to stellar mass was higher for the flatter systems. They interpreted this as evidence of rotational motion with varying projections along the line of sight, and they inferred a higher fraction of rotational support (as indicated by $V/\sigma$) for the high-$z$ systems compared to analogous local galaxies. 

An alternative approach is to directly measure rotation by resolving the stellar continuum in high-resolution spectra of gravitationally lensed objects. In a pilot project for the present study, \citet{Newman15} presented the first such measurements. They showed that the lensed massive quiescent galaxy MRG-M0150 at $z=2.64$ is rotating at $V \sin i = 189 \pm 34$~km~s${}^{-1}$,  a surprisingly high speed considering that its likely descendants are mostly ``slow rotators'' with $V/\sigma$ values that are far lower. \citet{Toft17} showed that the lensed massive quiescent galaxy M2129-1 (MRG-M2129 in our nomenclature) is a rotationally-supported disk galaxy and inferred an extremely high rotation speed of $V = 532^{+67}_{-49}$~km~s${}^{-1}$. These initial observations supported the idea that early quiescent galaxies frequently have disk-like morphologies and kinematics.

In Paper~I (Newman et al.~2018) we presented an imaging survey and associated follow-up observations with which we identified a sample of five quiescent galaxies at $z=1.95$-2.64 that are magnified by galaxy clusters. In four cases, the images are at least several arcseconds in extent and so can be resolved from the ground in good seeing conditions. In this paper we present the stellar kinematics of these four galaxies based on deep near-infrared (NIR) spectroscopic data. This sample includes MRG-M0150 ($z=2.64$), which was the subject of the pilot study by \citet{Newman15}, and MRG-M2129 ($z=2.15$), which was studied by \citet{Geier13} and \citet{Toft17}, along with two newly discovered lensed galaxies, MRG-M0138 ($z=1.95$) and MRG-P0918 ($z=2.36$). In the case of MRG-M2129, we have obtained independent observations that we compare to \citet{Toft17}.

In Section 2 we describe the resolved stellar kinematic measurements. In Section 3 we perform dynamical modeling of the three galaxies for which a lens model exists. In Section 4 we discuss our results in the context of low- and high-redshift samples and consider their implications for the formation and evolution of massive galaxies. Throughout we assume a flat $\Lambda$CDM cosmology with $\Omega_m = 0.3$ and $H_0=70$~km~s${}^{-1}$~Mpc${}^{-1}$.

\section{Spatially Resolved Stellar Kinematics}

In this section we present our measurements of the spatially resolved stellar kinematics of the lensed quiescent galaxies MRG-M0138, MRG-M0150, MRG-P0918, and MRG-M2129. As discussed in Paper~I, these are massive quiescent galaxies ($M_* \gtrsim 10^{11.0} \msol$ in cases with estimated magnification) with ages spanning the range 0.5-1.4~Gyr. Our survey also uncovered a fifth lensed quiescent galaxy, MRG-S1522, but we omitted it from this analysis because our spectrum is not well resolved. For MRG-M0150, we adopt the kinematics measured by \citet{Newman15}, which are reproduced in Table~\ref{tab:kin}. (We have verified that using the procedures described in this section, which differ only in detail from those of \citealt{Newman15}, results in consistent measurements.)

\subsection{Observations}

\begin{figure*}
\centering
\includegraphics[width=0.75\linewidth]{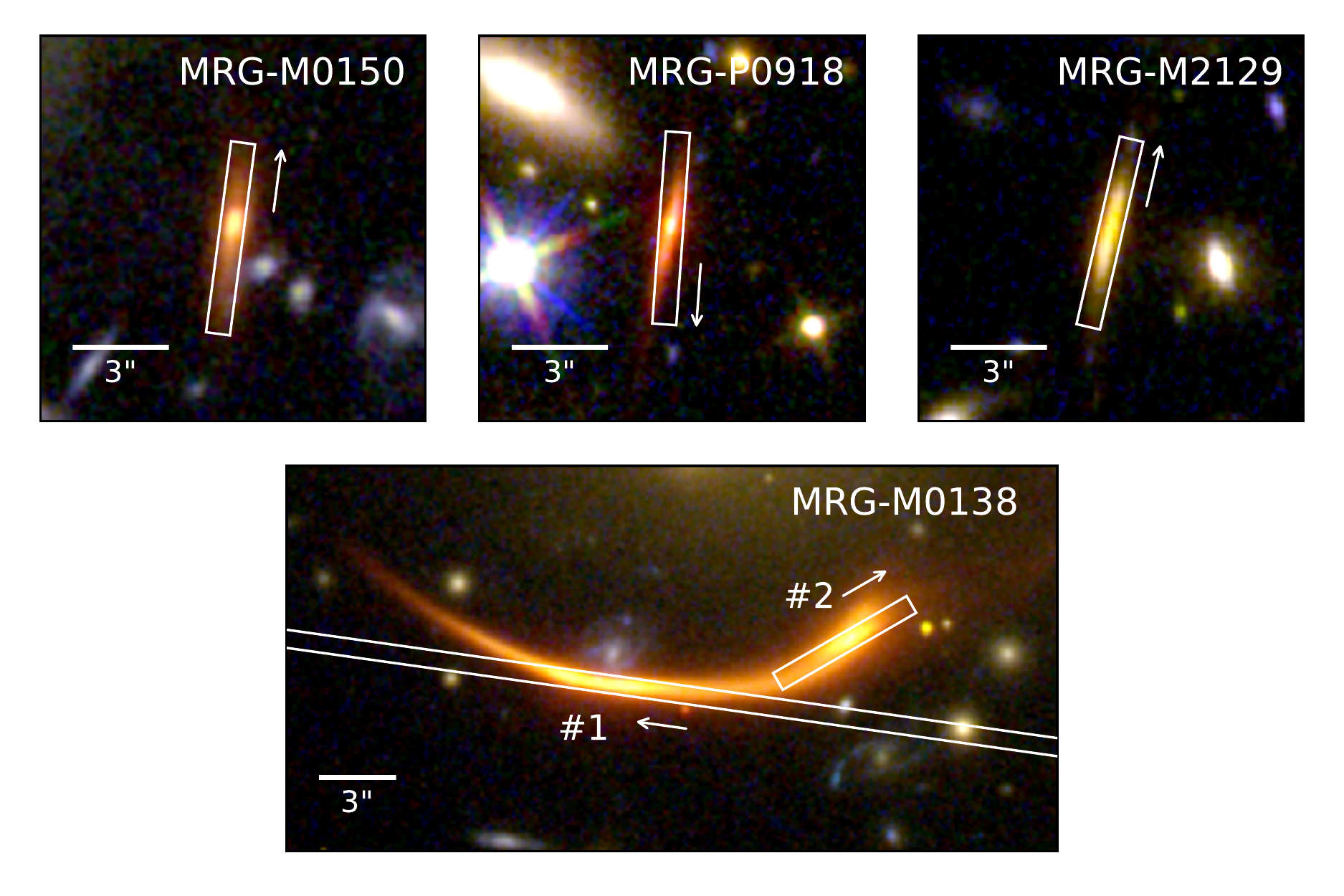}
\caption{Placement of slits on the lensed galaxies. The slit dimensions and orientations match the MOSFIRE (MRG-M0138) and FIRE (all others) observations. The images are shown centered in the slits, but the telescope was dithered between two positions during the observations. Arrows indicate the positive direction in the slit coordinate system used in Table~\ref{tab:kin} and Figure~\ref{fig:imagepl}. Observations were made of both Image 1 and 2 of MRG-M0138 (see text), but we only use those of Image~1 in this paper. North is up and east is left.\label{fig:slits}}
\end{figure*}

The near-infrared spectroscopic data and their reduction were described in Paper~I. Briefly, we used the FIRE echellette spectrograph \citep{Simcoe13} on the Magellan Baade telescope to observe MRG-M0150, MRG-P0918, and MRG-M2129 over the full NIR wavelength range. We also used MOSFIRE \citep{McLean12} at the Keck 1 telescope to observe MRG-M0138 Image 1 in the $J$ and $H$ bands and MRG-M0150 Image 1 in the $H$ band. The mean seeing for each target ranged from $0\farcs42$-$0\farcs79$, with a median of $0\farcs57$ (see Paper~I, Table~3). We note that FIRE observations were also undertaken for MRG-M0138 Image 2, as mentioned Paper~I, but since the slit orientation is close to the minor axis for this image, it is not well suited to measure rotation; we rely on the MOSFIRE observations of Image 1 instead.

\subsection{Stellar Kinematic Measurement Technique\label{sec:kinmethod}} 

\begin{deluxetable}{cccc}
\tablecolumns{4}
\tablewidth{0pt}
\tablecaption{Resolved Stellar Kinematic Measurements\label{tab:kin}}
\tablehead{\colhead{Bin lower} & \colhead{Bin upper} & \colhead{$v$ (km s${}^{-1}$)} & \colhead{$\sigma$ (km s${}^{-1}$)} \\
\colhead{edge (arcsec)} & \colhead{edge (arcsec)} & \colhead{} & \colhead{}}
\startdata
\cutinhead{MRG-M0138 Image 1}
-4.95 & -2.97 & $203 \pm 18$ & $227 \pm 19 \pm 11$\\
-2.97 & -2.25 & $181 \pm 19$ & $263 \pm 17 \pm 13$\\
-2.25 & -1.53 & $128 \pm 25$ & $268 \pm 22 \pm 13$\\
-1.53 & -0.81 & $147 \pm 20$ & $305 \pm 23 \pm 15$\\
-0.81 & -0.27 & $64 \pm 22$ & $311 \pm 32 \pm 16$\\
-0.27 & 0.27 & $-30 \pm 25$ & $326 \pm 20 \pm 16$\\
0.27 & 0.81 & $-129 \pm 26$ & $325 \pm 28 \pm 16$\\
0.81 & 1.71 & $-164 \pm 26$ & $248 \pm 31 \pm 12$\\
\cutinhead{MRG-M0150}
-0.60 & 0.60 & \ldots & $271 \pm 18 \pm 38$\\
-1.80 & -0.90 & $101 \pm 45$ & \ldots\\
-0.90 & -0.30 & $99 \pm 23$ & \ldots\\
-0.30 & 0.30 & $15 \pm 17$ & \ldots\\
0.30 & 0.80 & $-96 \pm 22$ & \ldots\\
0.80 & 1.40 & $-206 \pm 45$ & \ldots\\
\cutinhead{MRG-P0918}
-1.00 & -0.60 & $89 \pm 30$ & \ldots\\
-0.60 & -0.20 & $85 \pm 13$ & $189 \pm 18 \pm 21$\\
-0.20 & 0.20 & $0 \pm 13$ & $253 \pm 17 \pm 28$\\
0.20 & 0.60 & $-49 \pm 17$ & $198 \pm 21 \pm 22$\\
0.60 & 1.00 & $-55 \pm 28$ & \ldots\\
\cutinhead{MRG-M2129}
-1.40 & -0.75 & $-190 \pm 18$ & $196 \pm 28 \pm 14$\\
-0.75 & -0.25 & $-193 \pm 15$ & $195 \pm 16 \pm 14$\\
-0.25 & 0.25 & $10 \pm 25$ & $227 \pm 30 \pm 16$\\
0.25 & 0.75 & $181 \pm 24$ & $239 \pm 37 \pm 17$\\
0.75 & 1.40 & $245 \pm 24$ & $200 \pm 42 \pm 14$
\enddata
\tablecomments{The errors listed for $\sigma$ are the random and systematic components, respectively, and the latter are correlated within each galaxy. Bin limits are in arcsec along the slit relative to the peak flux, with the positive direction indicated by arrows in Figure~\ref{fig:slits}. The velocity zeropoint is determined from the dynamical model fits, except for MRG-P0918, where we set the central bin to zero.}
\end{deluxetable}

For each lensed galaxy, we extracted spectra in a series of bins along the slit. The orientations of the slits are shown in Figure~\ref{fig:slits}. (The slit was slightly misaligned for MRG-P0918, since the orientation of the image was initially estimated from a shallower ground-based image; we will discuss the small effects of this misalignment below.) The total spatial extent of the bins was determined by examining the intensity profile along the slit and selecting the region that exceeded $\simeq 15\%$ of the peak. Within this region, the size of the bins was approximately matched to the mean seeing during the observation, although in some cases the outer bins were enlarged to increase the signal-to-noise ratio (S/N). This procedure produced spectra with ${\rm S/N} \approx 10$-80 per 300~km~s${}^{-1}$ (approximately one velocity dispersion element) which is adequate to measure velocities and, in most of the bins, velocity dispersions.

We used the spectral range from $\lambda_{\rm rest} = 3600$~\AA~to the end of the $H$ band for the kinematic measurements. We omitted the $K$ band spectra since it does not contain strong absorption features besides H$\alpha$, which is contaminated by emission. A mask was created for each spectrum to exclude spectral regions that (1) contain strong telluric absorption bands, (2) contain emission lines, or (3) deviate from the stellar population model in ways that are likely to arise from data reduction problems (e.g., a small region blueward of H$\beta$ for MRG-M2129, and the 4300~\AA~region of MRG-M0138 where residual telluric absorption is evident). A small number of outlier pixels were identified via a $\sigma$-clipping algorithm and masked.

In the case of MRG-M0138, the MOSFIRE $J$ band spectrum begins around the Ca K line, so the continuum blueward of 4000~\AA~is not well constrained. For this galaxy we omitted the short spectral region blueward of 4000~\AA. We also masked Mg~b and Na~D, since these lines show clearly non-solar abundances.

\begin{figure*}
\centering
\includegraphics[width=0.45\linewidth]{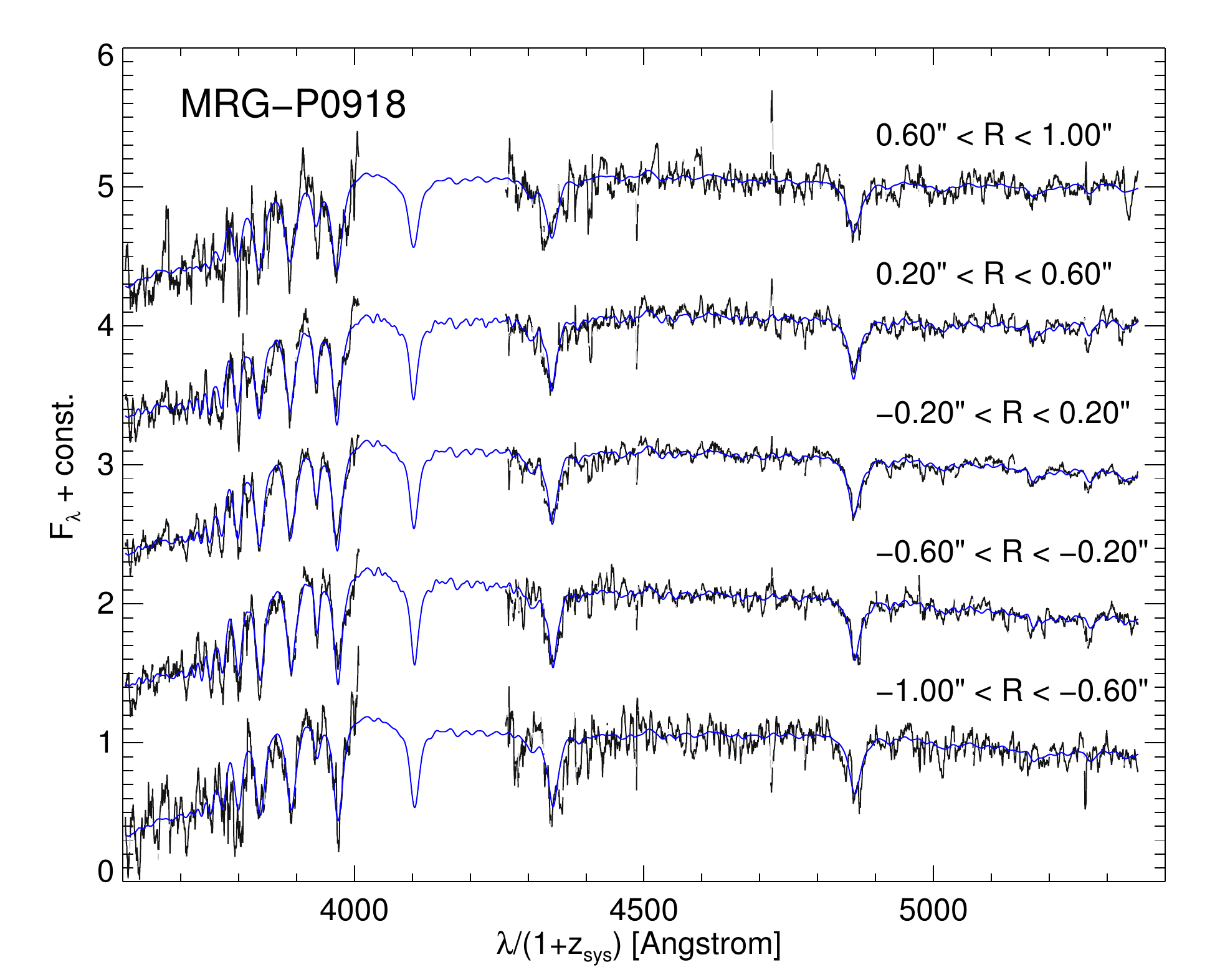}
\includegraphics[width=0.45\linewidth]{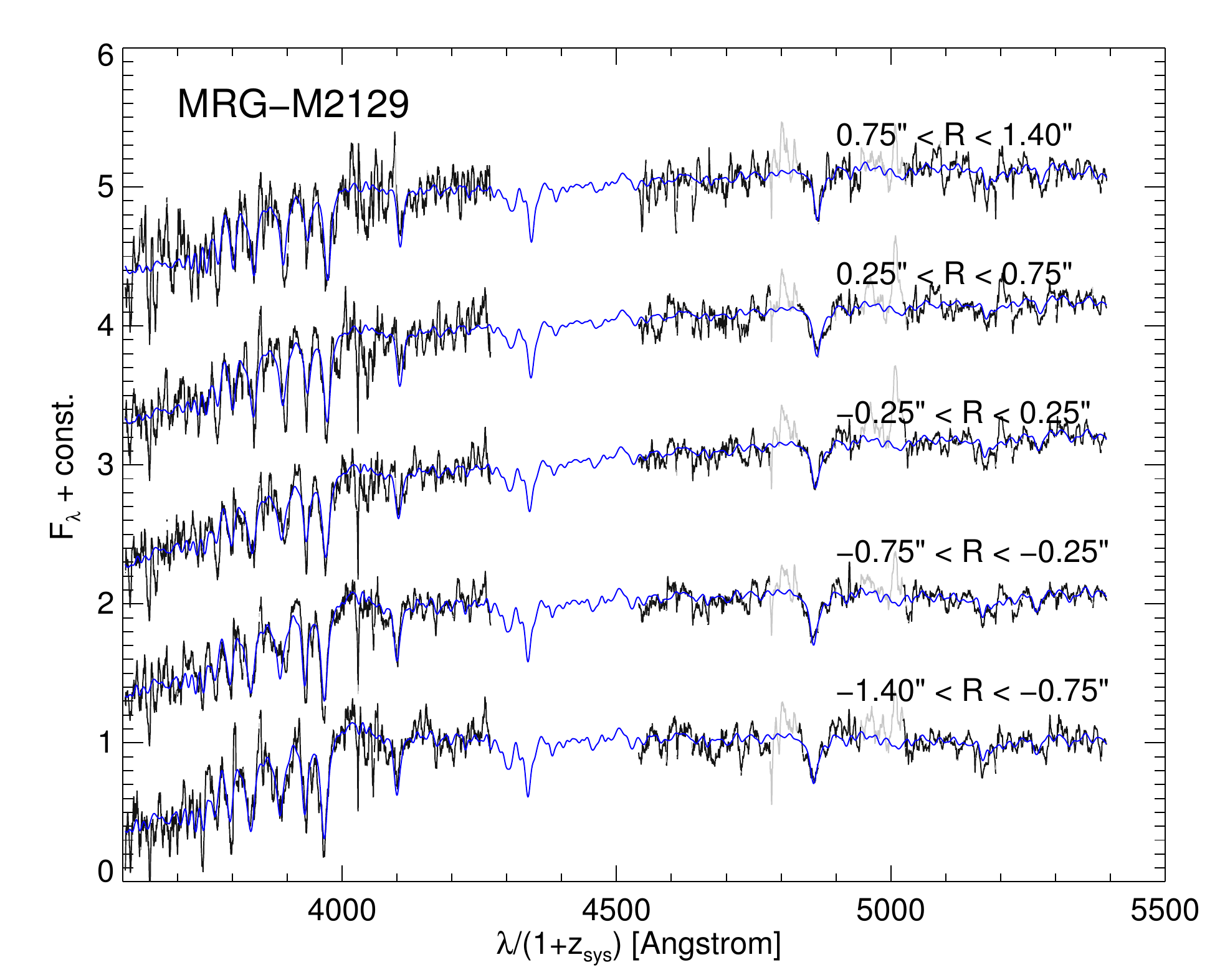}\\
\includegraphics[width=0.45\linewidth]{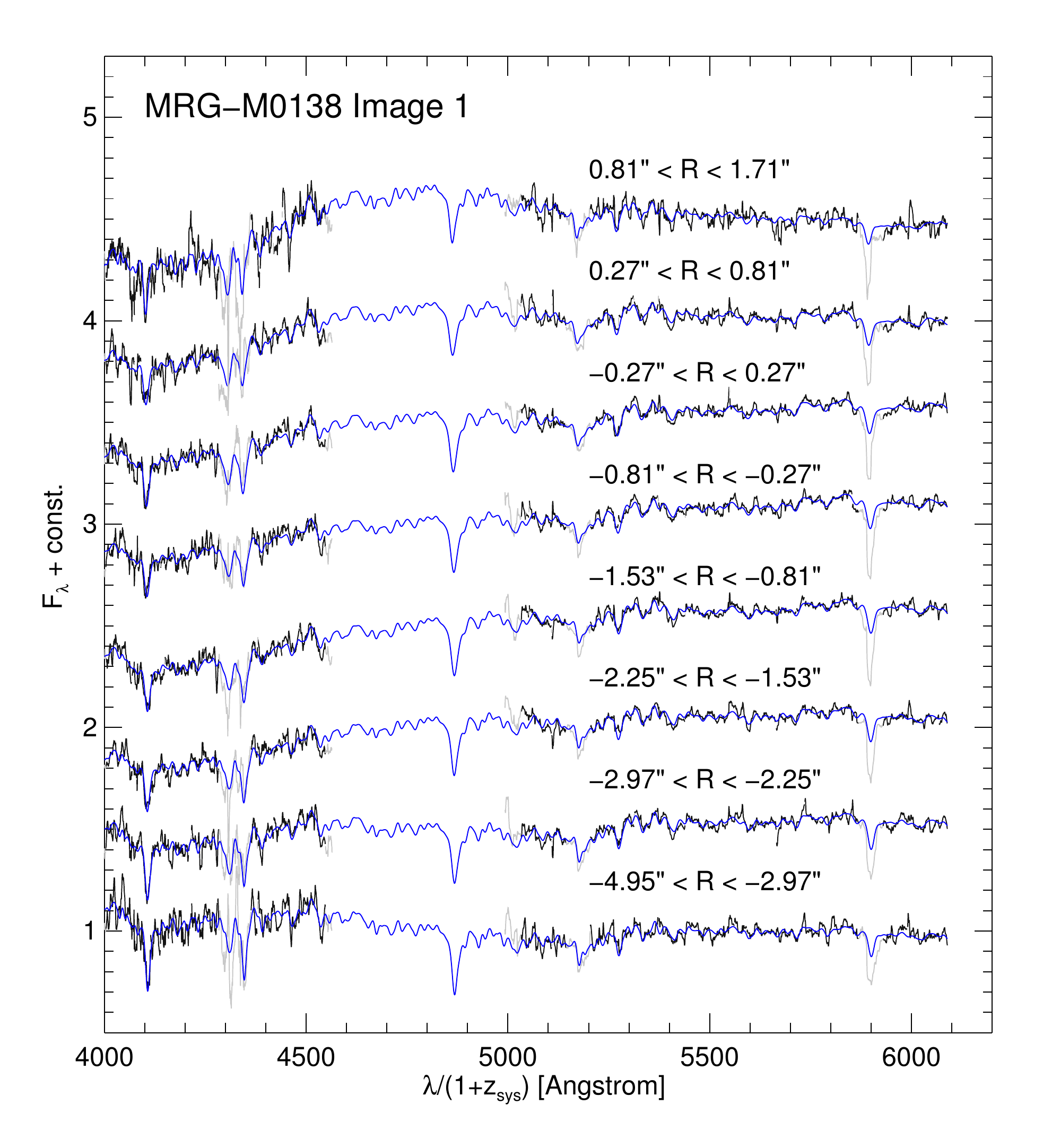}
\caption{Resolved spectra (grey) and model fits (blue) used to extract kinematics. Extraction apertures are specified in arcsec along the slit relative to the position of peak flux. Regions of the spectrum that were masked in the fit are shown in lighter grey. For display purposes, all spectra were smoothed with a $\simeq 390$~km~s${}^{-1}$ boxcar using inverse variance weighting.\label{fig:specmontage}}
\end{figure*}

In each bin, we used {\tt ppxf} \citep{Cappellari04} to measure the velocity $V$ and velocity dispersion $\sigma$. By default, the template spectrum was constructed by {\tt ppxf} as a linear combination of \citet{Vazdekis15} simple stellar populations with ages ranging up to the age of the universe at the observed epoch and solar-scaled abundances with metallicities ranging from approximately solar to twice solar. This template was broadened by a Gaussian line-of-sight velocity distribution, redshifted, multiplied by a linear polynomial, and added to a polynomial of degree $N$ to best fit the observed spectrum. By default we set $N \approx \Delta \lambda / (200~{\rm\AA})$, where $\Delta \lambda$ is rest-frame length of the portion of the spectrum used in the fit. Uncertainties were derived by shuffling the residuals in chunks to maintain correlations, adding these to the best-fit model, refitting a large number of such realizations, and measuring the scatter in $V$ and $\sigma$. The resulting uncertainty estimates were moderately larger than the formal uncertainties derived from the $\chi^2$ surface. Figure~\ref{fig:specmontage} shows the spatially resolved spectra and the model fits used to extract kinematics (see \citealt{Newman15} for MRG-M0150). Table~\ref{tab:kin} lists the stellar kinematic measurements.

We then varied our procedure in several ways in order to estimate the systematic uncertainties and assess the robustness of the measurements. First, we used solar-metallicity \citet{BC03} models instead of the \citet{Vazdekis15} grid. Second, we fit {\tt FSPS} \citep{Conroy09,Conroy10} models using the {\tt pyspecfit} \citep{Newman14} code instead of {\tt ppxf}, as described in Paper~I. Third, we varied the additive polynomial order by $\pm 2$. Fourth, we masked the Balmer lines, since these are sensitive to the star formation history and to the rotational velocities of the library stars. 

We found that the $\sigma$ measurements in the outer bins of MRG-P0918 were unstable to these changes and so omitted them from our analysis. \citet{Newman15} determined that only an integrated velocity dispersion could be robustly measured for MRG-M0150, and we adopt the $\sigma$ measured in that paper within a $\pm 0\farcs6$ aperture. The resolved velocity dispersions for MRG-M0138 and MRG-M2129 were stable in all of the spatial bins. 

For the stable bins, we found that varying the measurement procedure has a minimal effect of $\lesssim 3\%$ on the relative velocities, which are very robust. The main effect is to induce systematic shifts in $\sigma$. We quantified these as $\langle \Delta \sigma / \sigma_{\rm def} \rangle$; here the mean is taken over all spatial bins within a galaxy, and $\sigma_{\rm def}$ is the measurement derived with the default procedure. The largest shifts and their origins were $\langle \Delta \sigma / \sigma_{\rm def} \rangle = -5$\% for MRG-M0138 (BC03 templates), $-11$\% for MRG-P0918 (BC03 templates), and +7\% for MRG-M2129 (masking Balmer lines, or FSPS templates). For reference, \citet{Newman15} found a systematic uncertainty of 14\% for MRG-M0150. We will treat the absolute values of these shifts as perfectly correlated fractional systematic uncertainties within each galaxy when we perform dynamical modeling in Section~\ref{sec:modeling}. As expected, the spectra most dominated by Balmer absorption (MRG-M0150 and MRG-P0918) have the largest systematic uncertainties in $\sigma$, whereas the oldest system (MRG-M0138) is the most robust.

\subsection{Model-Independent Limits on Rotation\label{sec:modelindep}}

The resolved stellar kinematics are plotted in Figure~\ref{fig:imagepl}. All of the galaxies are clearly rotating. Figure~\ref{fig:sourcepl} shows the position of the slit in the source plane. For MRG-M0138, MRG-M0150, and MRG-M2129, the slit is approximately aligned along the major axis; this is by design for the multiply imaged cases (MRG-M0138 and MRG-M0150) and is fortuitous for MRG-M2129. (Recall that we lack a lens model for the fourth system, MRG-P0918.) Therefore we expect the long-slit kinematics to capture the majority of the projected rotation, but because of seeing and the width of the slit in the source plane, these data cannot be directly interpreted as major axis kinematics. That requires modeling of the velocity field (Section~\ref{sec:modeling}).

From the raw velocity measurements we can estimate only lower limits on the rotation speed, but we can do so in a model-independent way as half of the velocity difference between the outermost bins: $V_{\rm max}^{\rm proj} > 183 \pm 16$~km~s${}^{-1}$ for MRG-M0138 Image 1, $V_{\rm max}^{\rm proj} > 155 \pm 32$~km~s${}^{-1}$ for MRG-M0150, $V_{\rm max}^{\rm proj} > 72 \pm 21$~km~s${}^{-1}$ for MRG-P0918, and $V_{\rm max}^{\rm proj} > 218 \pm 15$~km~s${}^{-1}$ for MRG-M2129. The notation emphasizes that these are projected velocities, i.e., $V \sin i$, that are not corrected for inclination.

The position of the FIRE slit was slightly misaligned with the MRG-P0918 image. Likewise the MRG-M0138 Image 1 is curved and so is unavoidably miscentered in the slit for most of its length. This will lead to spurious shifts in the measured velocities. For each of these galaxies, we calculated the miscentering of the flux in the slit for each spatial bin using the \emph{HST} image, and we converted these to velocity shifts based on the spectrograph parameters. The velocity shifts are $\lesssim 6$~km~s${}^{-1}$ in all cases. Since these are much smaller than the measurement uncertainties, they will be neglected for the remainder of the paper.

\section{Dynamical Modeling}
\label{sec:modeling}
\begin{figure*}
\includegraphics[width=0.47\linewidth]{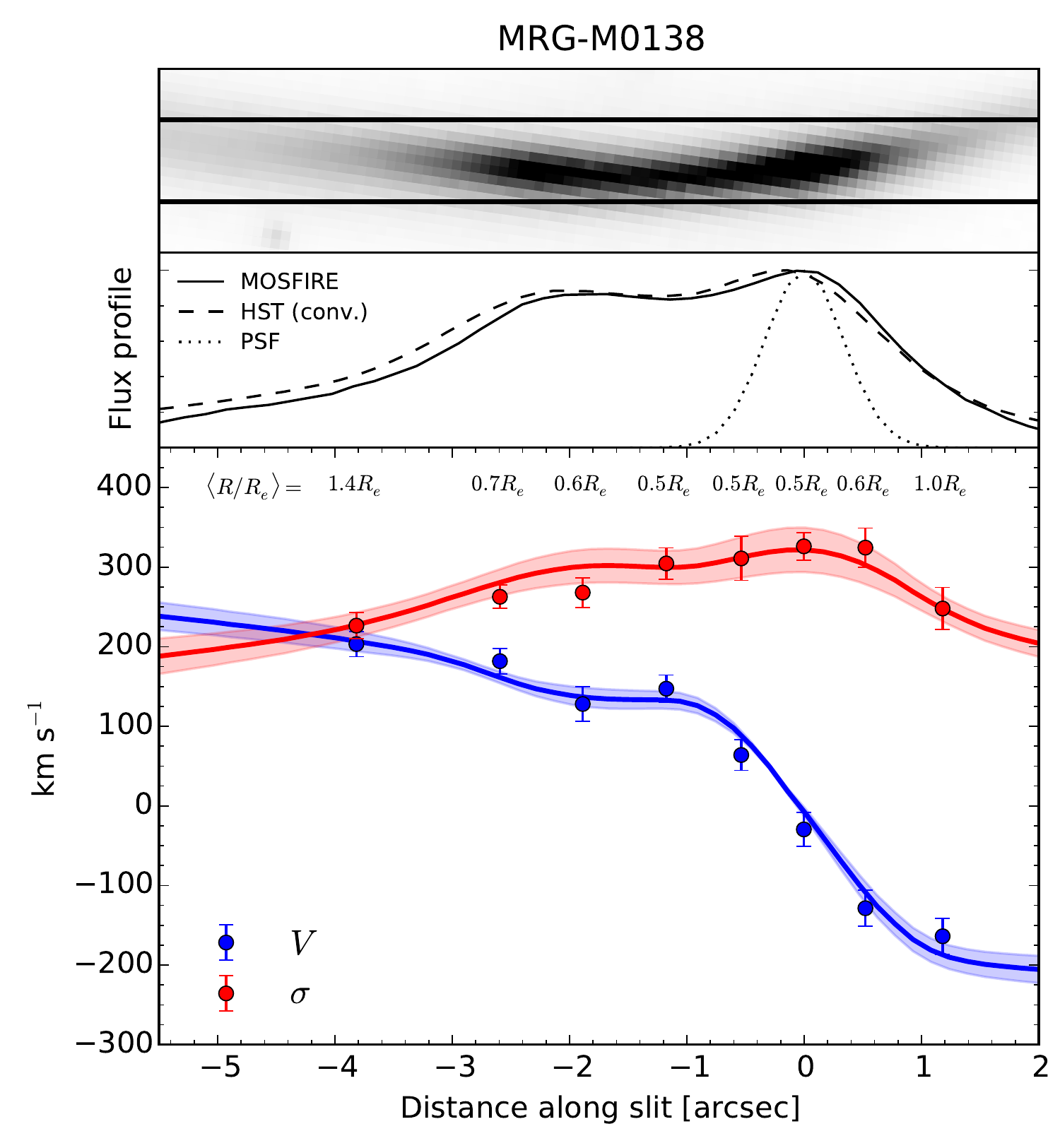}\hfill
\includegraphics[width=0.47\linewidth]{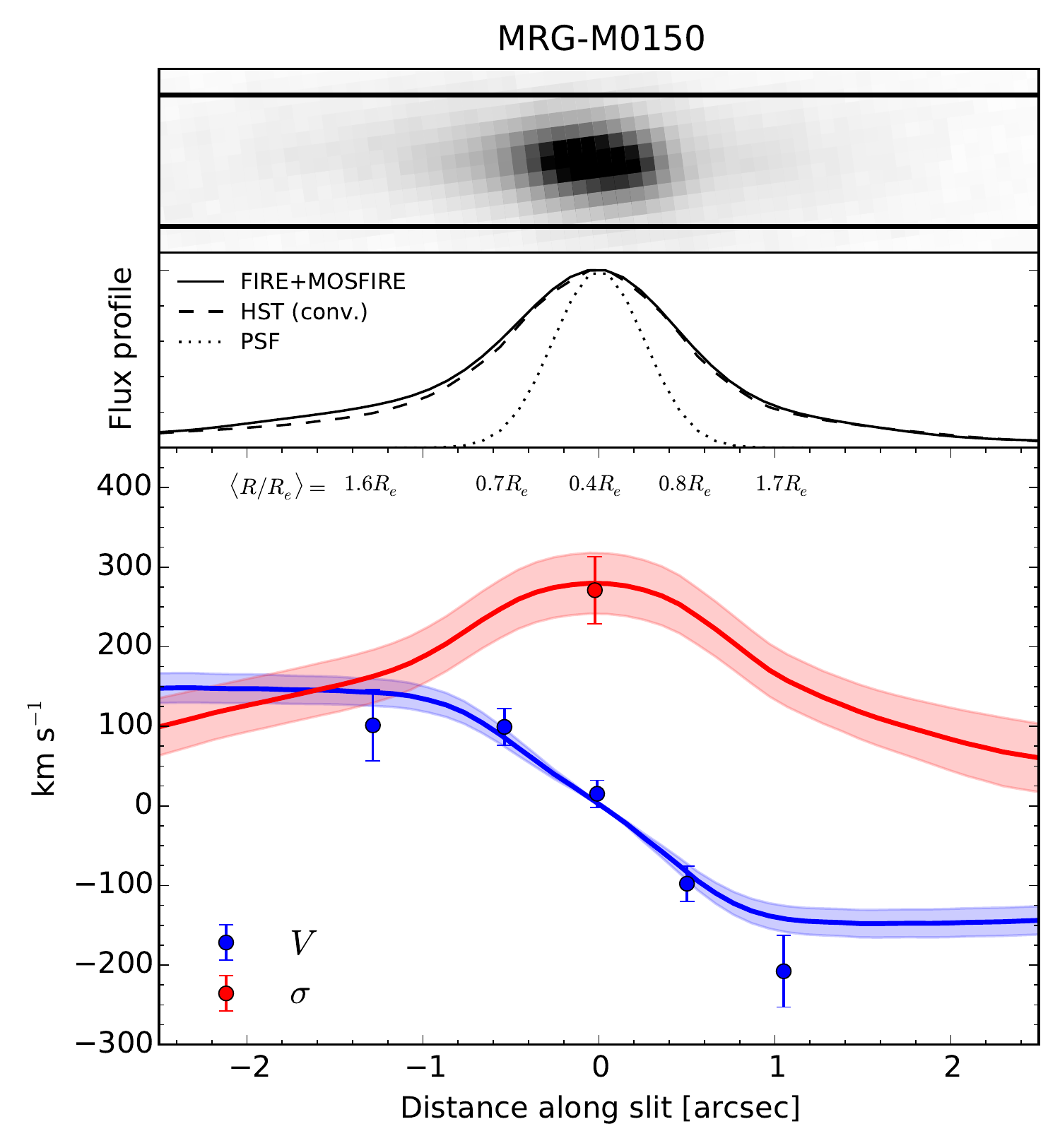}\\
\includegraphics[width=0.47\linewidth]{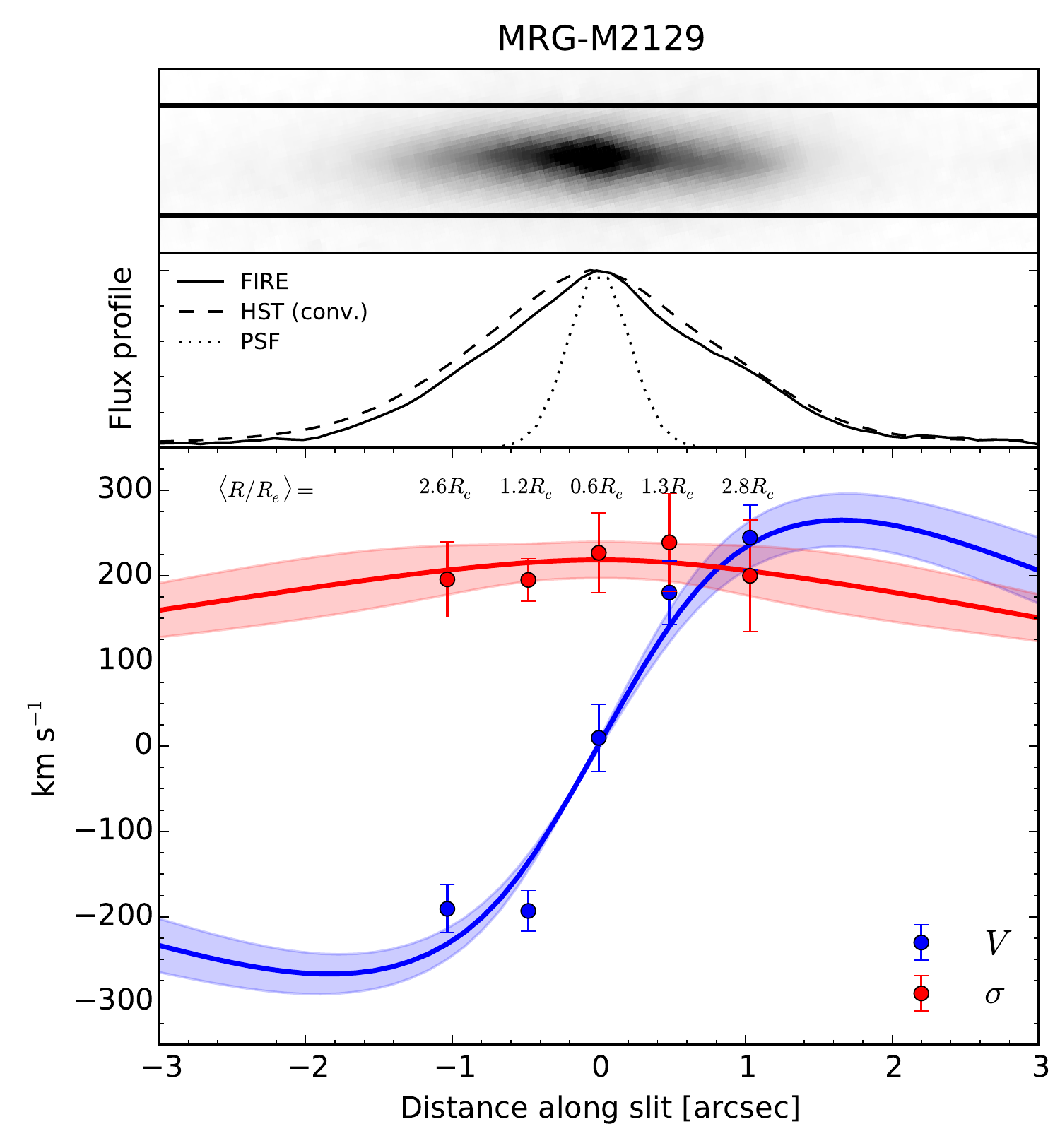}\hfill
\includegraphics[width=0.47\linewidth]{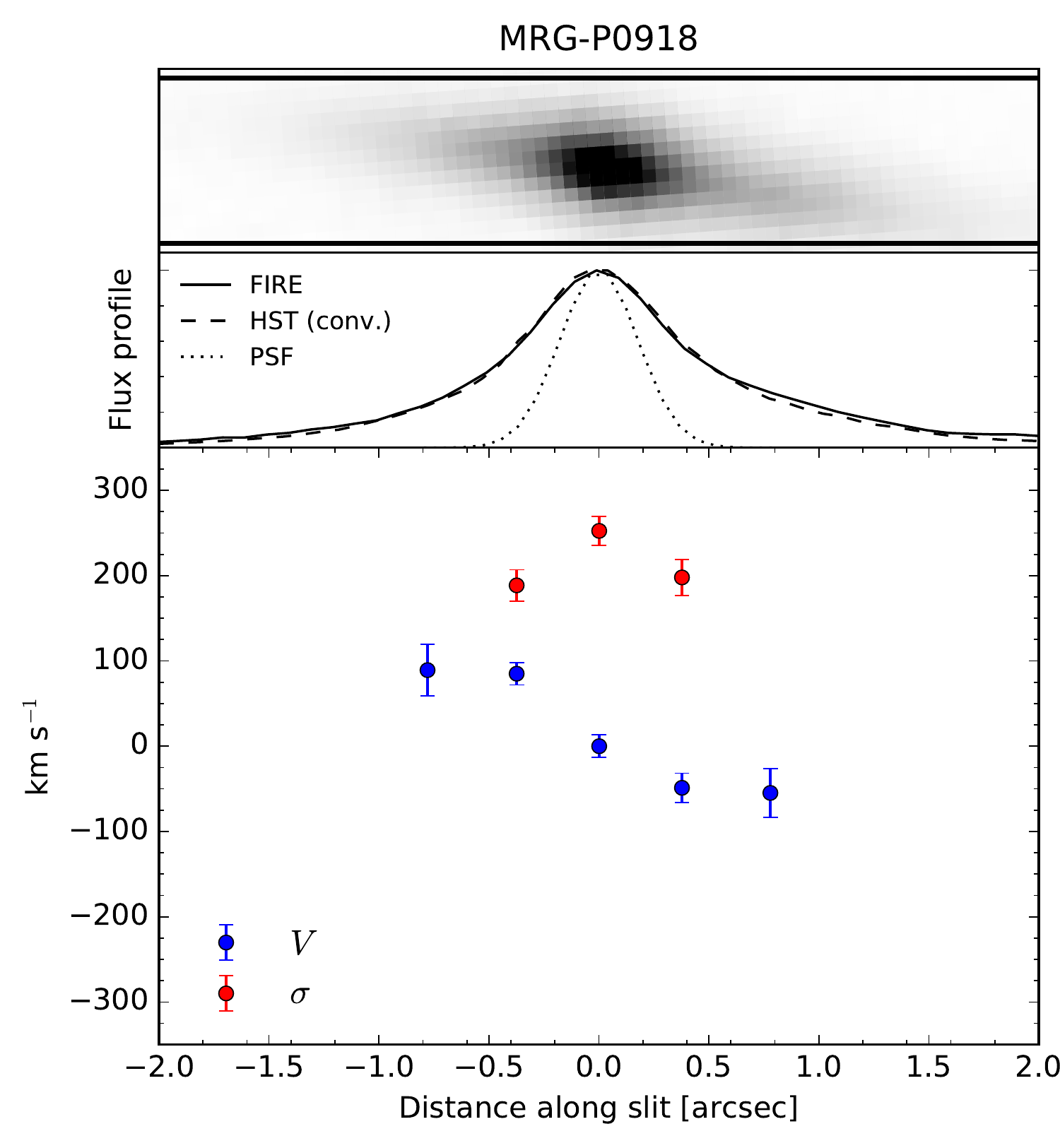}
\caption{Stellar kinematic data and dynamical model in the image plane. For each galaxy, the lower panel shows the measured velocities and velocity dispersions along with the model, which takes into account the lens mapping and observational effects (binning and blurring by seeing). The colored lines are the mean of models drawn from the posterior, and the width of the bands indicates their standard deviation. The flux-weighted radius represented in the spectrum in each spatial bin is indicated as $\langle R/R_e\rangle$. Measurement errors have been rescaled by $m_{\rm err}$ (see text). The middle panels show the flux profile observed in the spectrum (solid line) and in the \emph{HST} F160W image after blurring by the seeing and integrating across the slit width (dashed line). The PSF is also indicated to demonstrate that the spectra are spatially resolved. The upper panels show the \emph{HST} F160W image (not convolved) with a linear stretch and the position of the slit overlaid.\label{fig:imagepl}}
\end{figure*}

\begin{figure*}
\includegraphics[width=0.47\linewidth]{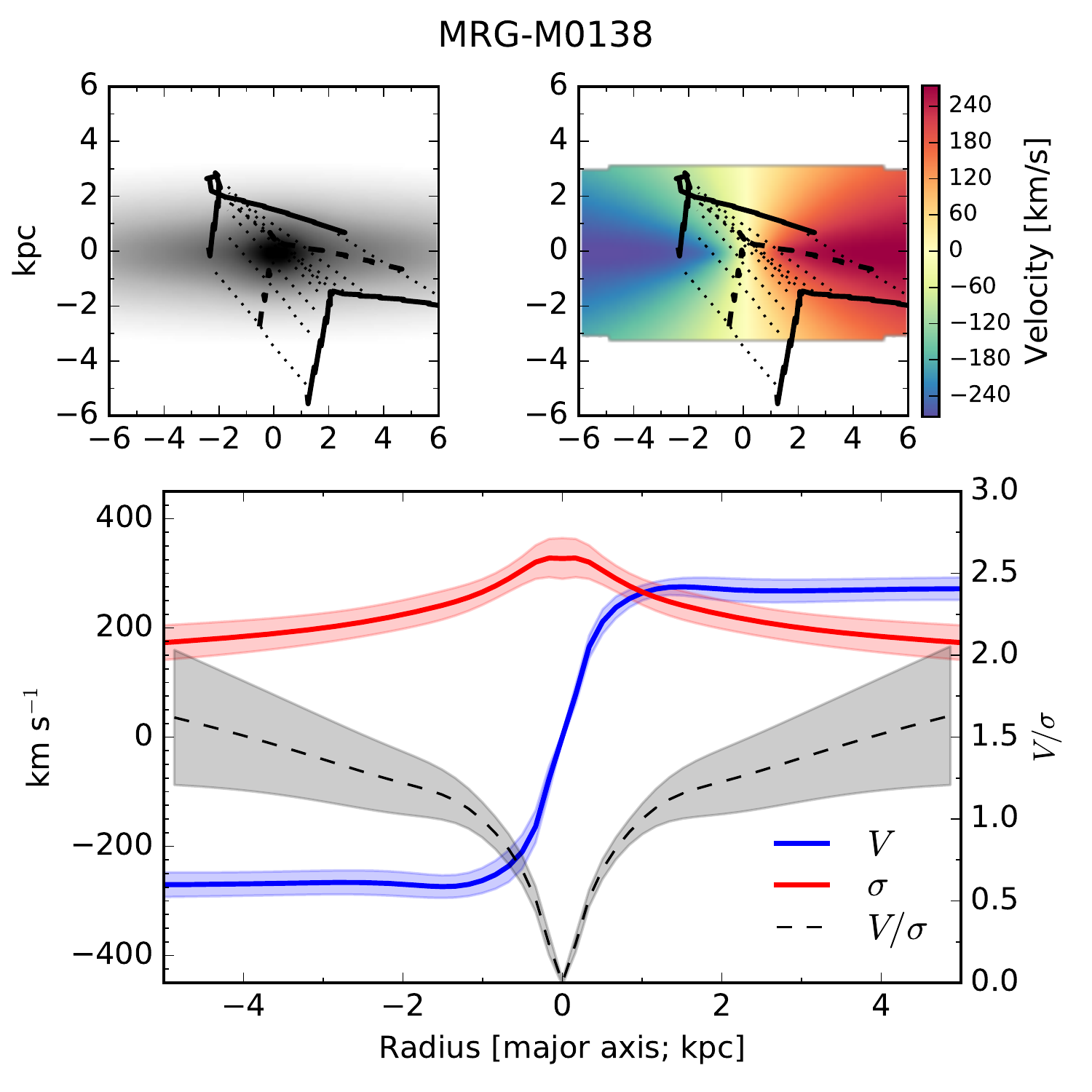}\hfill
\includegraphics[width=0.47\linewidth]{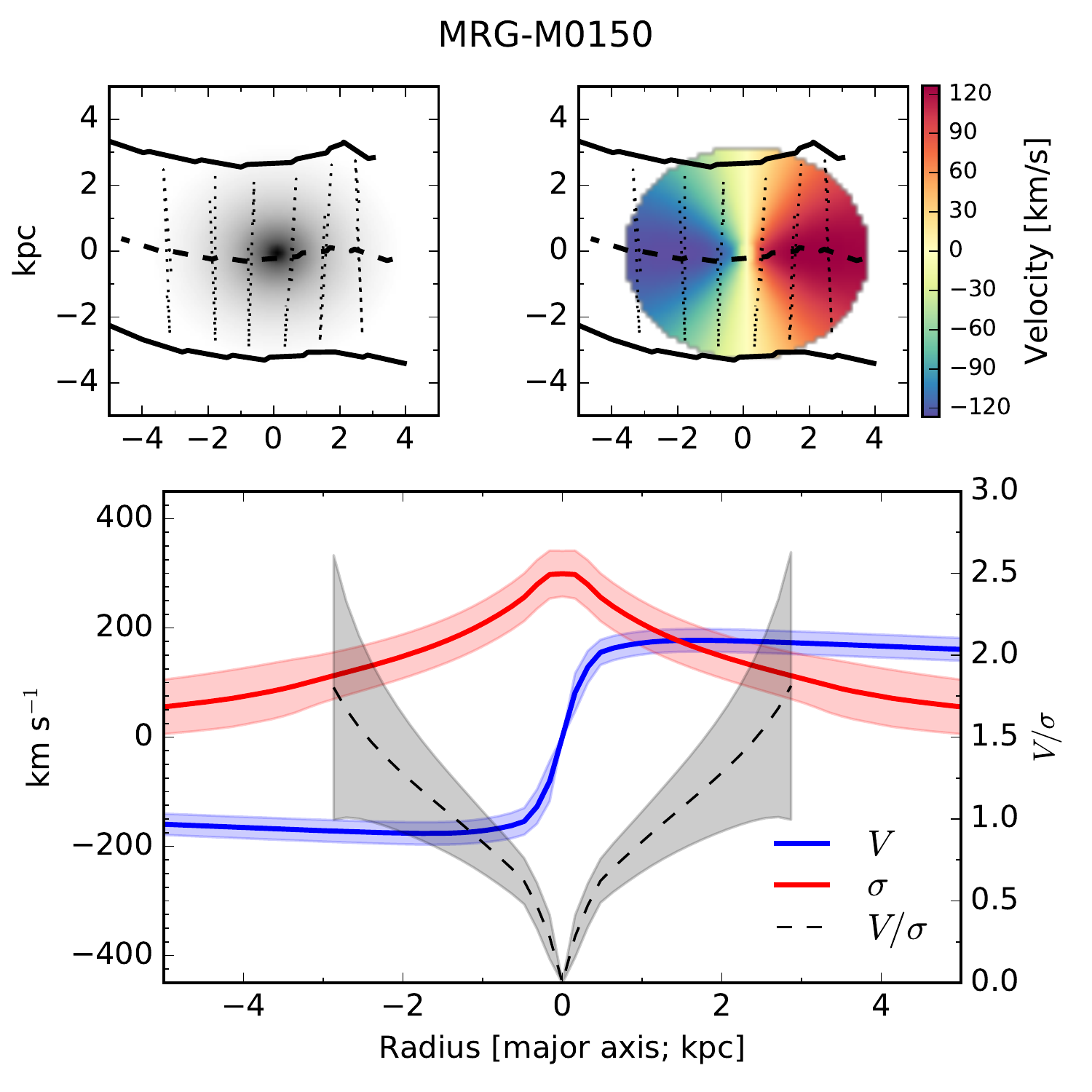}\\
\includegraphics[width=0.47\linewidth]{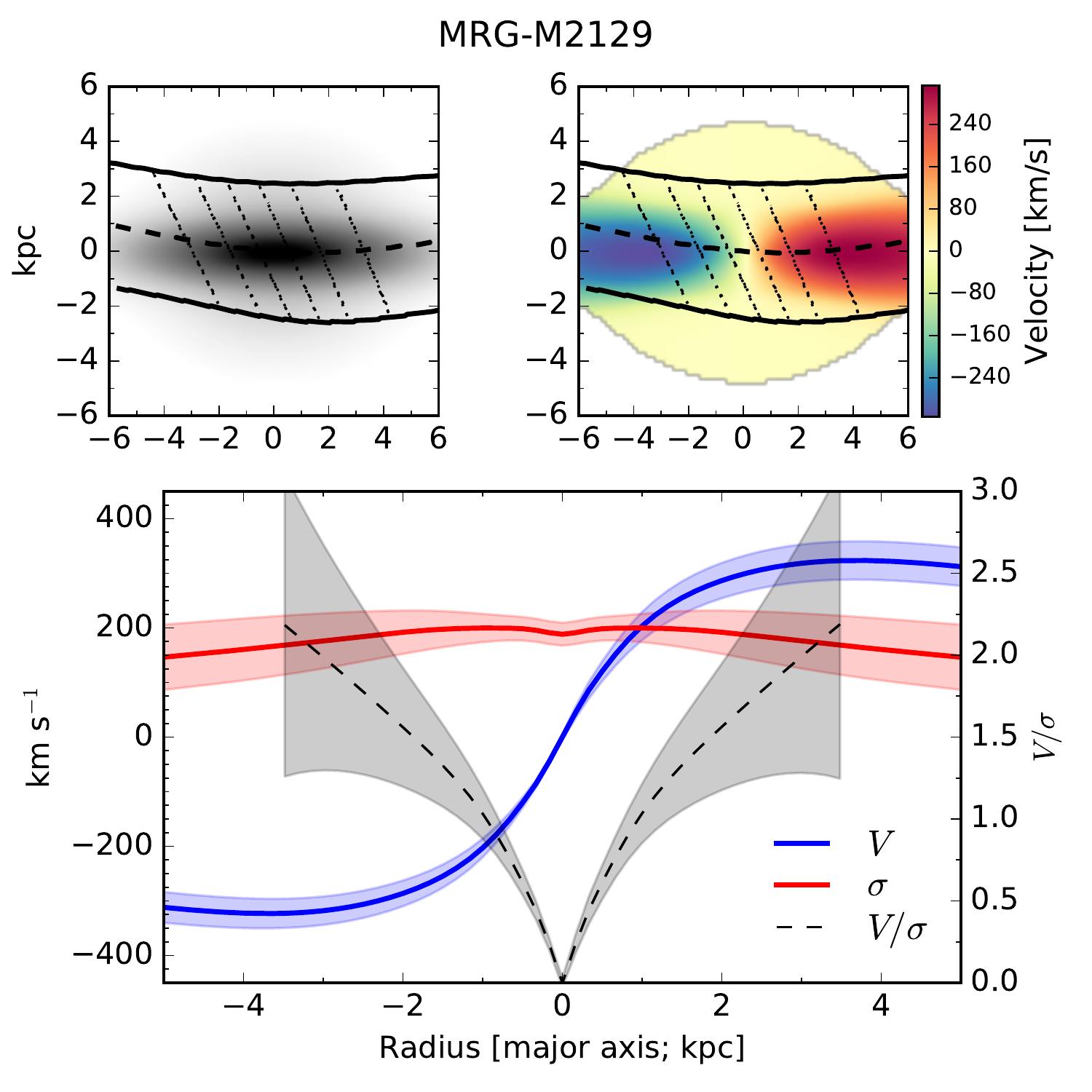}\hfill
\caption{Stellar dynamical models in the source plane. For each galaxy with a lens model (i.e., excluding MRG-P0918), the lower panel shows the projected velocity and velocity dispersion along the major axis. The local projected $V/\sigma$ ratio is plotted in gray out to the radius where its $1\sigma$ uncertainty is $\pm 1$. The widths of the bands indicate the standard deviation of models drawn from the posterior. The upper panels show the source surface brightness (upper left panels) with a logarithmic stretch and the velocity field (upper right panels). The spectrograph slit edges mapped to the source plane are shown as solid lines, with the middle of the slit drawn as a dashed line. The dotted lines denote the boundaries of the bins in which the spectra were extracted. (The small loop formed by one slit edge for MRG-M0138 occurs because it crosses the caustic.)
\label{fig:sourcepl}}
\end{figure*}

Dynamical modeling is required not only to measure masses, but also to model the velocity field and estimate the maximum rotation velocity on the major axis $V_{\rm max}^{\rm proj}$ and associated quantities such as $V/\sigma$. These cannot be directly estimated from the observations because of several effects: (1) the galaxy is resolved only along the direction of maximum magnification, i.e., along the slit, (2) this direction is not necessarily aligned with the major axis, (3) the slit width over which the kinematics are integrated therefore has a large physical extent, and (4) seeing blurs the images.

We developed a model to compute stellar kinematics and trace these through the lens mapping, accounting for the seeing and binning of the observations. The model is described by three sets of parameters that relate to the galaxy structure, the stellar kinematics, and the measurement errors. We will now describe each in turn.

The first set of parameters describes the single- or double-S\'{e}rsic model of the source (i.e., effective radius $R_e$, S\'{e}rsic index $n$, and axis ratio $q=b/a$ for each S\'{e}rsic component, along with the relative luminosities in the case of two components). These structural parameters are well constrained by the source plane reconstructions of the \emph{HST} images presented in Paper~I (Table~4). We nevertheless used Gaussian priors in order to propagate their uncertainties.\footnote{In cases where two uncertainties are listed in Table~4 of Paper~I, we use only the first to define this Gaussian prior; the second uncertainty reflects systematic uncertainties in the magnification factor, which are treated below.}

The kinematic parameters are the dynamical mass $\log M_{\rm dyn}$, the velocity anisotropy $\beta_z$, the dimensionless parameter $\kappa$ that describes rotation (see below), and the inclination angle $i$. To compute the stellar kinematics, we used the Jeans Anisotropic Modeling (JAM) methods developed by \citet{Cappellari08}. The models assume that the galaxies are oblate, that light traces mass, and that the velocity ellipsoid is aligned with a cylindrical coordinate system and has a shape parameterized by $\beta_z = 1 - \langle v_z^2 \rangle/\langle v_R^2 \rangle$. We assumed that $\beta_z$ is spatially uniform and adopted a broad uniform prior $\beta_z \sim U(0, 0.5)$ based on the range seen in local early-type galaxies \citep{Cappellari07}. Since we cannot assume that all of the galaxies are thin disks, the inclination $i$ is not uniquely determined from the ellipticity. We took a uniform prior on $\cos i \sim U(0, \cos i_{\rm max})$, where $\cos i_{\rm max}$ was defined to enforce that all S\'{e}rsic components have an intrinsic ellipticity $e_{\rm intr} < 0.85$.

Given the surface brightness distribution and values of $i$ and $\beta_z$, the projected second moments $\langle v^2 \rangle$  are determined up to a constant factor that is set by the dynamical mass $M_{\rm dyn}$. Velocities are calculated following \citet[][see also \citealt{Cappellari08}]{Satoh80}. In this approach, which has proved useful for analyzing low-redshift galaxies, the azimuthal motion $\langle v_{\phi}^2 \rangle$ is partitioned into streaming $\langle v_{\phi} \rangle$ and dispersion $\sigma_{\phi}$ components as $\langle v_{\phi} \rangle = \kappa [ \langle v_{\phi}^2 \rangle - \langle v_R^2 \rangle ]^{1/2}$. Therefore $\kappa$ is a dimensionless parameter that specifies the amplitude of rotation relative to the special case in which $\sigma_{\phi} = \sigma_R$ ($\kappa = 1$; see Section~\ref{sec:rotspeeds}). Both $M_{\rm dyn}$ and $\kappa$ are well constrained by our data. These data do not have the resolution needed to constrain $i$ and $\beta_z$, which affect the detailed structure of the second moment map, but we include these parameters in order to marginalize over them.

The third set of parameters, $m_{\sigma}$ and $m_{\rm err}$, relate to the measurement uncertainties. We defined a multiplicative factor $m_{\sigma}$ that scales the velocity dispersions and thereby accounts for the correlated systematic uncertainties in $\sigma$. We adopted a Gaussian prior on $m_{\sigma}$ that is centered on unity and has a dispersion taken from Section~\ref{sec:kinmethod}. A second factor $m_{\rm err}$ was used to scale the measurement errors. We took a Gaussian prior on $\ln m_{\rm err}$ centered on 0 with a dispersion of 0.7, i.e., a factor of 2. This method enlarges the measurement uncertainties and widens the posterior distributions if required to fit the data. As discussed below, we ultimately found that the measurement errors required little or no rescaling.

\begin{figure*}
\centering
\includegraphics[width=0.7\linewidth]{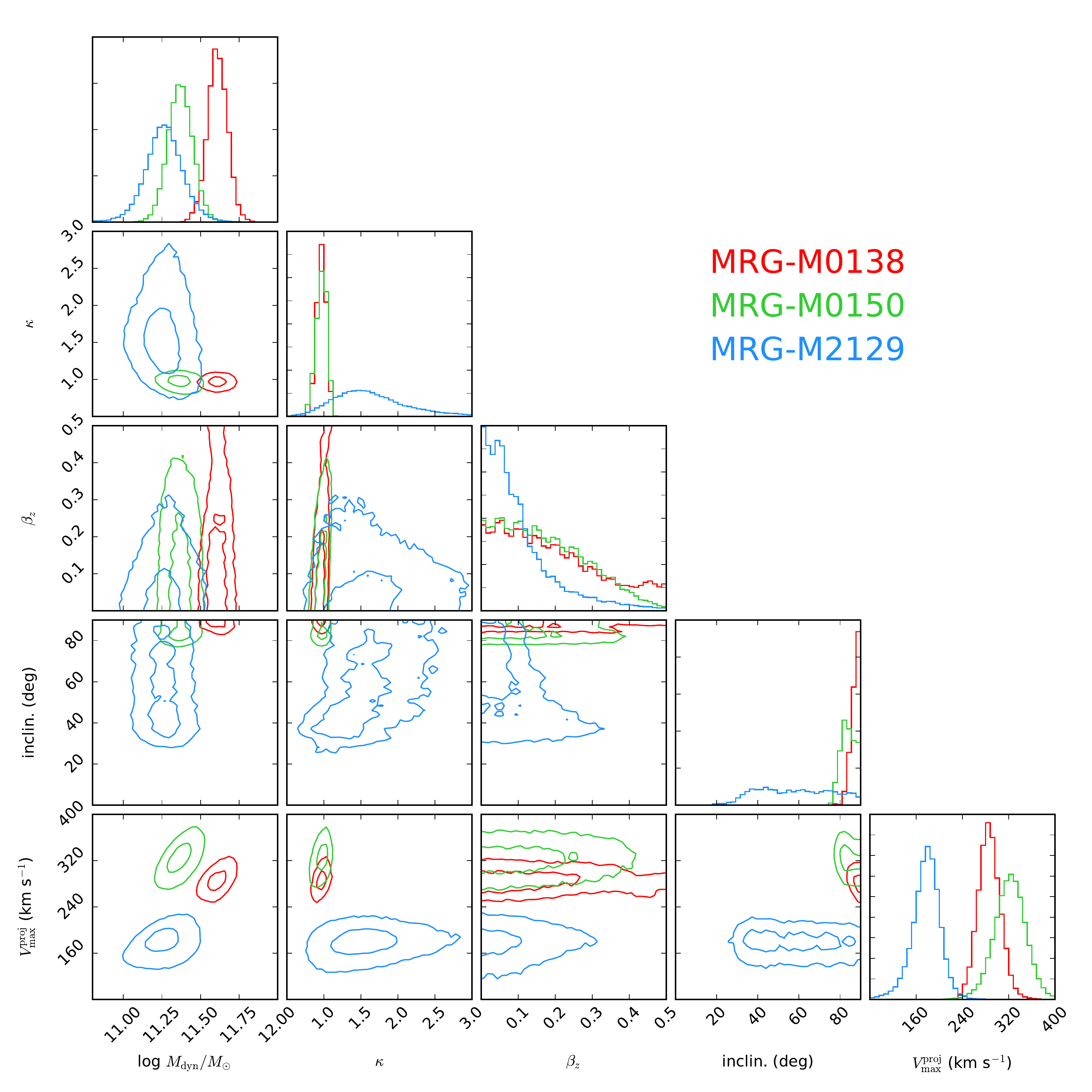}
\caption{Covariances in stellar dynamical models for the three lensed quiescent galaxies identified in the legend. For each galaxy, contours enclose 68\% and 95\% of the posterior probability. Note that $V^{\rm proj}_{\rm max}$ is not an independent parameter but follows from $M_{\rm dyn}$ and $\kappa$. This figure was prepared using {\tt corner.py} \citep{cornerpy}.\label{fig:corner}}
\end{figure*}

For a given set of parameters, we calculated the likelihood $L$ as follows. First, we generated the surface brightness distribution of the source and fit it with a multi-Gaussian expansion, as required by the JAM routines. We then used JAM to compute the projected stellar kinematics on a fine grid in the source and image planes. The lens models described in Paper~I define the mapping between the two planes. The image plane was then convolved by a Gaussian PSF and binned to match the observations. For each spatial bin $j$, this results in a model velocity $V^{\rm mod}_j$ and a second moment $\langle V^2_j \rangle$ projected along the line of sight. We computed the model velocity dispersion $\sigma^{\rm mod}_j$ as $\langle V^2_j\rangle = (V^{\rm mod}_j)^2 + (\sigma^{\rm mod}_j)^2$. The likelihood is then
\begin{multline}
L(\log M_{\rm dyn}, \kappa, \beta_z, i, \Delta V_{\rm sys},m_{\sigma}, m_{\rm err}, \rm{Sersic\ parameters}) \\
 = \prod_{j=1}^N \frac{1}{\sqrt{2 \pi} m_{\rm err} \sigma_{v_j}} \exp \left[ -\frac{1}{2} \left( \frac{V_j - \Delta V_{\rm sys} - V^{\rm mod}_j}{m_{\rm err} \sigma_{v_j}} \right)^2 \right] \\
\times \prod_{k=1}^M \frac{1}{\sqrt{2 \pi} m_{\rm err} \sigma_{\sigma_k}} \exp \left[ -\frac{1}{2} \left( \frac{\sigma_k - m_{\sigma} \sigma^{\rm mod}_k}{m_{\rm err} \sigma_{\sigma_k}} \right)^2 \right]
\label{eqn:chi2}
\end{multline}
Here $V_j$ and $\sigma_k$ are the measured velocities and velocity dispersions, respectively, with associated uncertainties $\sigma_{V_j}$ and $\sigma_{\sigma_k}$. Shifts to the galaxy systemic velocity $\Delta V_{\rm sys}$ are small and precisely determined, so we sped the calculation by fixing the value of $\Delta V_{\rm sys}$ that maximizes $L$ at the values of the other parameters. The data in Table~\ref{tab:kin} have been shifted by the $\Delta V_{\rm sys}$ values determined from the model fits.

Although the model contains 10-14 parameters (depending on whether the source has one or two S\'{e}rsic components), we emphasize that the kinematics are largely determined by the photometry and just two kinematic parameters: $\log M_{\rm dyn}$ and $\kappa$. The other parameters are included mainly to marginalize over and thereby propagate uncertainties.

We used Monte Carlo methods to marginalize over the posterior probability distributions and derive the constraints listed in Table~\ref{tab:kinresults}. Covariances among the key dynamical parameters are shown in Figure~\ref{fig:corner}. As expected, $i$ is not constrained beyond the prior, which imposes a narrow range on MRG-M0138 and MRG-M2129 near $90^{\circ}$. Similarly we find only loose constraints on $\beta_z$, so we do not list posterior constraints for these parameters in Table~\ref{tab:kinresults}. They are not strongly covariant with the main parameters of interest, which are well determined: $\log M_{\rm dyn}$, $\kappa$, and the projected quantities $V_{\rm max}^{\rm proj}$ and $\lambda_{R_e}$ which are defined below. The one exception is the mild covariance between $\kappa$ and $i$ for MRG-M0150, which we will discuss in Section~\ref{sec:rotspeeds}. 

Figures~\ref{fig:imagepl} and \ref{fig:sourcepl} show the fitted dynamical models in the image and source planes, respectively, and demonstrate their good fits to the data. The parameter $m_{\rm err}$ that rescales the measurement errors is consistent with unity for MRG-M0138 and MRG-M0150 (see Table~\ref{tab:kinresults}), while for MRG-M2129 only a modest increase in the errors by $\simeq 1.5\times$ is needed. 

\subsection{Lens Model Uncertainties}

In Paper~I, we estimated the uncertainties in the source structural parameters via two methods for the multiply imaged galaxies MRG-M0138 and MRG-M0150: the image-to-image scatter (which addresses the internal consistency of the lens model) and a magnification uncertainty (which addresses systematic uncertainties in the lens model). So far we have propagated the former component via the Gaussian priors described above, but we have not propagated the latter. This is more difficult because uncertainties in the lens model affect both the source structure and the mapping from the image plane kinematics to the source plane. As a simple first-order test of the effect of changing the magnification by a factor $x$, we scaled the source luminosity and $R_e$ by $x^{-1}$ and $x^{-1/2}$, respectively, and isotropically dilated the source plane coordinates by a factor $x^{-1/2}$. The only effect of changing the magnification by the uncertainties estimated in Paper~I was to shift $\log M_{\rm dyn}$ by 0.06~dex for MRG-M0138 and MRG-M0150. This is smaller than the fractional magnification uncertainty, which is expected since $M_{\rm dyn} \propto \sigma^2 R_e \propto \mu^{-1/2}$ to a first approximation. We have therefore added 0.06~dex in quadrature to the uncertainty in $\log M_{\rm dyn}$ for these systems and report these enlarged errors in Table~\ref{tab:kinresults}. For MRG-M2129 the magnification uncertainty is sufficiently small (10\%) that it does not contribute significantly to the error budget.

\subsection{Metrics of Rotational Support and Angular Momentum}

The velocity and velocity dispersion fields are completely specified by the parameters described above. However, it is also useful to express the results in terms of other derived parameters. We calculated three of these:

\begin{itemize}
\item $V_{\rm max}^{\rm proj}$ is the maximum projected velocity on the major axis. The deprojected rotation velocity is $V_{\rm max} = V_{\rm max}^{\rm proj} / \sin i$.
\item To compare to local early-type galaxies, we calculated $\lambda_{R_e} = \langle R V \rangle / \langle R \sqrt{\sigma^2+V^2} \rangle$, where the averaging is weighted by flux and restricted within $R_e$. This parameter is a proxy for the projected specific angular momentum; it was defined by \citet{Emsellem07,Emsellem11} and applied to the SAURON and ATLAS${}^{\rm 3D}$ surveys. \item To compare to disk galaxies, we define $(V/\sigma)_{R_e}$. Disk galaxy kinematics, particularly at high redshifts, are often quantified by the ratio $V/\sigma$, where $\sigma$ is assumed to be uniform and isotropic throughout the disk. Since $\sigma$ varies with radius in our JAM models, we define a characteristic value $(V/\sigma)_{R_e}$, where $V=V^{\rm proj}(R_e) / \sin i$ is deprojected, $\sigma = \sigma^{\rm proj}(R_e)$, and both quantities are evaluated on the major axis. We find that $V$ is nearly maximal at $R_e$ in our sample, so the main effect of this choice is to pin down a fiducial radius at which to measure $\sigma$. We note that this definition is intended to compare to disk kinematics and differs from the conventional use of the central $\sigma_0$ in local early-type galaxies and from the metric used by \citet{Newman15}.
\end{itemize}
Constraints on these parameters are listed in Table~\ref{tab:kinresults}. The deprojected quantities $V_{\rm max}$ and $(V/\sigma)_{R_e}$ rely on further assumptions that we will describe in Section~\ref{sec:results}.

\begin{deluxetable*}{lccccc|cc}
\tablecolumns{8}
\tablewidth{0pt}
\tablecaption{Stellar Dynamical Model Constraints\label{tab:kinresults}}
\tablehead{\colhead{Galaxy} & \colhead{$\log M_{\rm dyn}/\msol$} & \colhead{$\kappa$} & \colhead{$V_{\rm max}^{\rm proj} ({\rm km}~{\rm s}^{-1})$} & \colhead{$\lambda_{R_e}$} & \colhead{$\ln m_{\rm err}$} & \colhead{$V_{\rm max} ({\rm km}~{\rm s}^{-1})$} & \colhead{$(V/\sigma)_{R_e}$} }
\startdata
MRG-M0138 & $11.64 \pm 0.10$ & $0.94 \pm 0.07$ & $290 \pm 21$ & $0.69 \pm 0.04$ & $-0.14 \pm 0.20$ & $290 \pm 21$ & $1.62 \pm 0.42$\\
MRG-M0150 & $11.26 \pm 0.14$ & $1.59 \pm 0.47$ & $176 \pm 26$ & $0.41 \pm 0.08$ & $0.00 \pm 0.33$ & $352 \pm 87$ & $2.29 \pm 0.73$\\
MRG-M2129 & $11.38 \pm 0.08$ & $0.96 \pm 0.07$ & $323 \pm 28$ & $0.69 \pm 0.05$ & $0.45 \pm 0.24$ & $323 \pm 28$ & $1.74 \pm 0.50$
\enddata
\tablecomments{Quantities left of the vertical line are projected. Quantities to the right are corrected for inclination assuming $i \approx 30^{\circ}$ for MRG-M0150 and $i \approx 90^{\circ}$ for MRG-M0138 and MRG-M2129; see Section~\ref{sec:results}. The mean and standard deviation of the marginalized posteriors are shown. See text for a discussion of the uncertainties.}
\end{deluxetable*}

\begin{figure}
\centering
\includegraphics[width=0.85\linewidth]{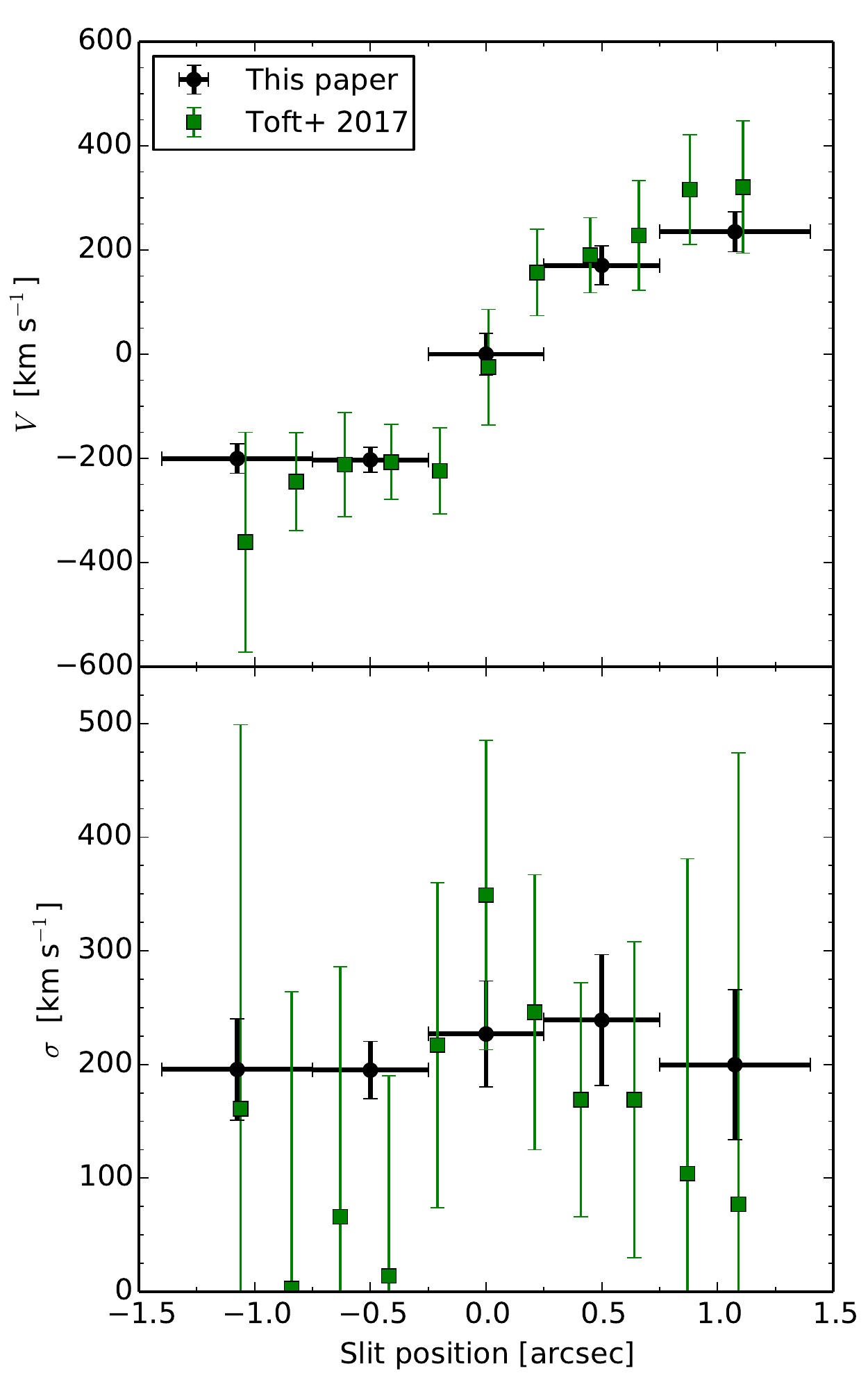}
\caption{Comparison of the observed stellar kinematics of MRG-M2129 measured in this paper (black points) and by \citet[][green points]{Toft17}.\label{fig:compare_toft}}
\end{figure}

\subsection{Comparison to Toft et al.}

\citet{Toft17} analyzed an X-Shooter spectrum of MRG-M2129 and inferred an extremely rapid rotation speed of $532^{+67}_{-49}$~km~s${}^{-1}$, which is 65\% higher than our less extreme value of $323 \pm 28$~km~s${}^{-1}$. To investigate the origin of this difference, in Figure~\ref{fig:compare_toft} we compare the observed kinematics. The Toft et al.~velocities are $\simeq 30\%$ higher in the outer bins, although they are still consistent within the uncertainties. Our dynamical models differ from those used by Toft et al.~in several respects, but the most important is the inclination angle. As discussed in the Appendix of Paper~I, Toft et al.~find a rounder source with an intermediate inclination of $i = 54^{\circ}$, whereas our reconstructed source has significantly higher projected ellipticity and so must have $i \approx 90^{\circ}$.\footnote{Given that the ellipticities of the flatter components of MRG-M0138 and MRG-M2129 are $b/a = 0.19$ and 0.24, respectively, under our assumption that $e_{\rm intr} < 0.85$ we must have $i > 80^{\circ}$ and $\sin i > 0.98$, justifying the assumption that $\sin i \approx 1$.} After a discrepancy in their treatment of the point spread function (PSF) was corrected, Toft et al.~found a higher ellipticity that now is lower than ours by only $\Delta e = 0.1$ (S.~Toft, A.~Man, et al., private communication). The difference between our inclination and the value published  by \citet{Toft17} translates to a factor of $\sin 54^{\circ} / \sin 90^{\circ} = 1.24$ in the deprojected rotation speed. Thus, about half of the total difference in deprojected rotation speed is attributable to differences in the kinematic measurements, while the other half arises from the inclination.  

The other important kinematic parameters are $M_{\rm dyn}$ and $V/\sigma$. We agree with Toft et al.~on the total dynamical mass to 0.1~dex. Toft et al.~find $V_{\rm max}/\sigma > 3.3$ at 97.5\% confidence. This parameter is not uniquely defined in our JAM models, since $\sigma$ is not constant. However, at $R_e$ we find a lower value of $(V/\sigma)_{R_e} = 1.74 \pm 0.50$ (see Table~\ref{tab:kin} and the radial variation in Figure~\ref{fig:sourcepl}). The cause of this difference can be seen in Figure~\ref{fig:compare_toft}: we measure $\sigma \approx 200$~km~s${}^{-1}$ across the image, in contrast to Toft et al.~who find much lower $\sigma$ in the outer bins. The Toft et al.~spectrum is shallower and the derived $\sigma$ measurements have much larger uncertainties, particularly in the outer bins where Toft et al.~only measure upper limits on $\sigma$. This might explain the difference. Based on their $V/\sigma$, Toft et al.~claim that MRG-M2129 is as rotationally supported as local late-type disk galaxies. Our lower value makes this less likely, although as we will show below, it is still much more rotationally supported than local early-type galaxies.

\begin{figure*}
\centering
\includegraphics[width=0.7\linewidth]{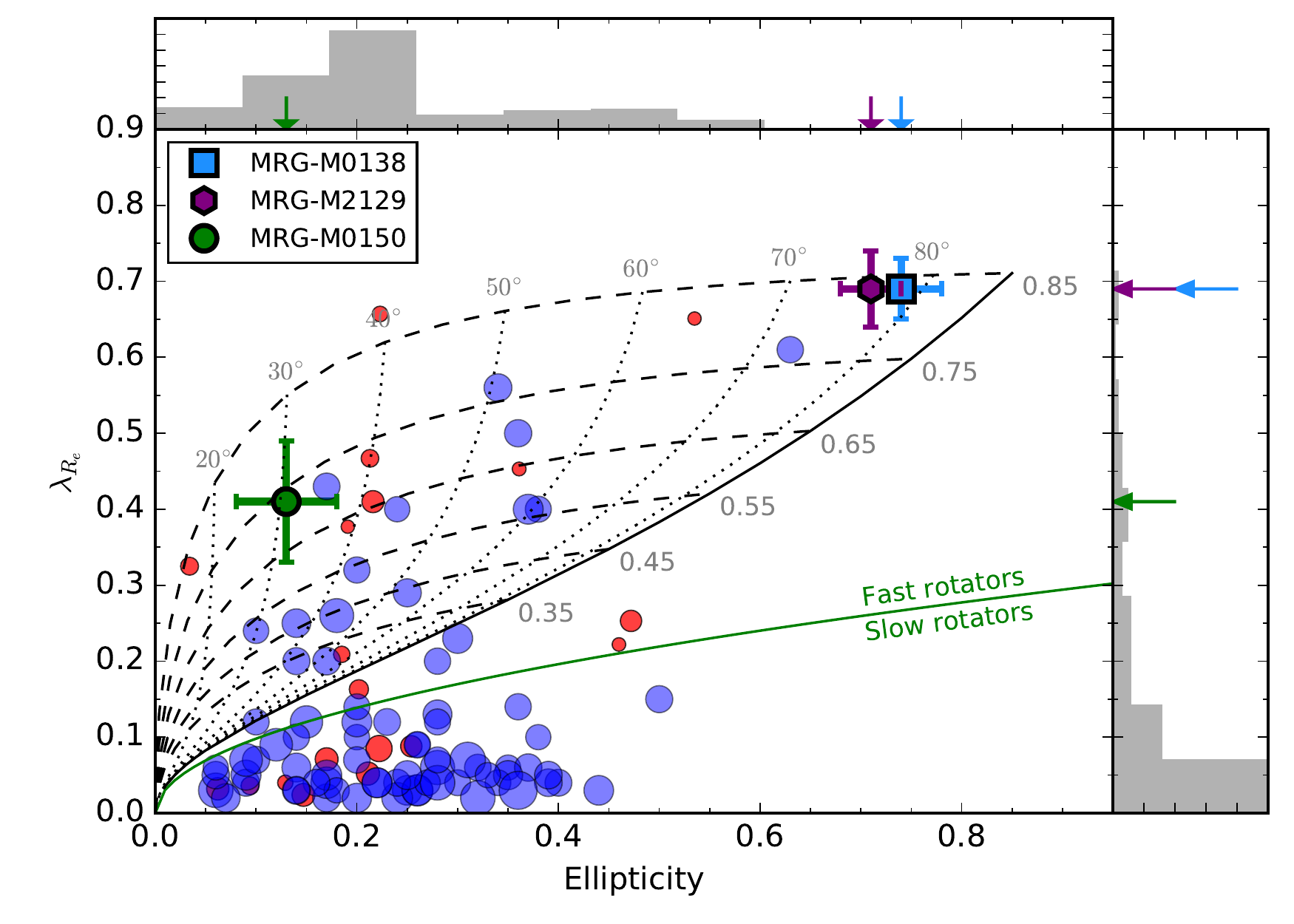}
\caption{The projected ellipticities and angular momentum parameters $\lambda_{R_e}$ of the $z=1.95$-2.64 lensed quiescent galaxies (points with error bars) are compared to local early-type galaxies from the \ATD~\citep[red circles,][]{Emsellem11,Cappellari13} and MASSIVE surveys \citep[blue circles,][]{Veale17}. Local galaxies with $\log M/\msol > 11.2$ are shown (see footnote 3). The symbol area is proportional to $\log M$ for the local systems. The grid is a family of models with $\beta_z = 0.65 e_{\rm intr}$ as found for local fast rotators \citep{Emsellem11}. Dashed lines have constant $e_{\rm intr}$ (labeled on the right) and dotted lines have constant inclination (labeled at top). The division between fast and slow rotators proposed by Emsellem et al.~is shown in green. Histograms show the \ATD+MASSIVE sample with weights applied based on the volumes of the two surveys, as described in Section~\ref{sec:disc_local}; the high-$z$ galaxies indicated by arrows. The figure shows that the lensed quiescent galaxies have similar flat intrinsic shapes ($e_{\rm intr} \approx 0.75$-0.85) and much more specific angular momentum than a typical early-type galaxy of equal or higher mass in the local universe.\label{fig:lambda}
}
\end{figure*}

\section{Results}
\label{sec:results}

\subsection{Rotation Speeds and Intrinsic Shapes}
\label{sec:rotspeeds} 

In general, determining the inclination of early-type galaxies is difficult because of their wide range of intrinsic shapes. For MRG-M0138 and MRG-M2129 the situation is unambiguous: since they have highly elliptical isophotes in projection, they must be intrinsically flat galaxies viewed nearly edge-on. MRG-M0150, however, is nearly round in projection ($b/a = 0.87$); based on photometry alone, we cannot tell whether it is nearly spherical or is a face-on flattened system. Kinematic data can distinguish these possibilities.

Figure~\ref{fig:lambda} shows the location of our lensed sample in the ellipticity versus $\lambda_{R_e}$ diagram. From integral field studies of local early-type galaxies, it has been found that the locus of ``fast rotators'' in this diagram can be described by simple models of single-component systems that span a range of intrinsic ellipticities and inclinations and have a velocity anisotropy $\beta_z$ that is proportional to the intrinsic ellipticity \citep{Emsellem11}. 

Under these assumptions, it is possible to estimate the shape and inclination of the galaxies in our sample by comparing them to the model grid in Figure~\ref{fig:lambda}. (We will compare our sample to the local early-type galaxies shown in the Figure in Section~\ref{sec:disc_local}). We infer that MRG-M0150 is an intrinsically flat galaxy ($e_{\rm intr}\approx 0.75$) viewed nearly face on ($i \approx 30^{\circ}$). The factor $1 / \sin i = 2.0 \pm 0.4$, leading to a deprojected rotation speed of $V_{\rm max} = 352 \pm 87$~km~s${}^{-1}$. For MRG-M0138 and MRG-M2129, we find $i > 80^{\circ}$ (see footnote 2) and can approximate $\sin i \approx 1$ so that $V_{\rm max} = 290 \pm 21$~km~s${}^{-1}$ and $323 \pm 28$~km~s${}^{-1}$, respectively. Although the relations describing $\beta_z$ in local early-type galaxies might not hold in $z > 2$ systems, we found that the inferred inclination of MRG-M0150 does not change much even if we compute the model grid in Figure~\ref{fig:lambda} under the very different assumption that $\beta_z = 0$. Alternatively, we can estimate $i$ by selecting \ATD~galaxies \citep{Emsellem11,Cappellari13} whose position in Figure~\ref{fig:lambda} is consistent with MRG-M0150. These have a median $i=38^{\circ}$, lower than the $57^{\circ}$ expected for randomly oriented disks and consistent with our model-based estimate.

Local early-type galaxies show a bimodal distribution of $\kappa$, with the ``fast rotators'' centered on 0.99 with a small rms of 0.07 \citep{Cappellari16}. MRG-M0138 and MRG-M2129 are consistent with $\kappa = 1$ within small uncertainties. They therefore have the simple dynamical structure that characterizes local fast rotators: the rotation speed follows from the galaxy shape and $\sigma_{\phi} = \sigma_R$, i.e., the velocity ellipsoid is oblate, so the rotational motion is just sufficient to produce the flattened galaxy shape without excess azimuthal dispersion $\sigma_{\phi} > \sigma_R$. For MRG-M0150 we find $\kappa = 1.59 \pm 0.47$ when marginalizing over inclination. Values of $\kappa > 1$ are physically possible but are somewhat unnatural; they imply more rotation than is needed to produce the flattened galaxy shape, which must be compensated by low $\sigma_{\phi}$. However, $\kappa$ is covariant with inclination (Figure~\ref{fig:corner}): if $i \approx 30^{\circ}$, as we argued above, then we find $\kappa = 1.20 \pm 0.30$ for MRG-M0150, closer to unity. This provides additional evidence that the three lensed galaxies have similar a dynamical structure viewed at different angles. It also supports the validity of the lens models and source plane reconstructions, since we would not expect $\kappa = 1$ if the ellipticity were seriously in error.

Although we cannot place MRG-P0918 in Figure~\ref{fig:lambda}, we can put a lower limit on $\lambda_{R_e}$. To so do, we derived a relation between $\lambda_{R_e}$ and $V^{\rm proj}_{\rm max} / \sigma_e$ within the \ATD~sample \citep{Emsellem11}, where $\sigma_e$ is the integrated velocity dispersion within $R_e$. We then used the lower limit $V^{\rm proj}_{\rm max} > 72$~km~s${}^{-1}$ from Section~\ref{sec:modelindep} and approximated $\sigma_e = 223$~km~s${}^{-1}$ using the integrated velocity dispersion measured in Paper~I. We find $\lambda_{R_e} \gtrsim 0.2$ for MRG-P0918. Although the division shown in Figure~\ref{fig:lambda} formally allows for a galaxy in this $\lambda_{R_e}$ range to be classified as a ``slow rotator'' if its ellipticity is sufficiently high, there are no such galaxies in the entire \ATD~sample. We therefore consider that MRG-P0918 can be classified as a ``fast rotator.''

In summary, all three of MRG-M0138, MRG-M0150, and MRG-M2129 are likely to be intrinsically flat, disk-dominated galaxies ($e_{\rm intr} \approx 0.75$-0.85) that are rotationally supported ($(V/\sigma)_{R_e} = 1.6$-2.3; Table~\ref{tab:kinresults}) and have rapid rotation speeds of $V_{\rm max}  = 290$-352~km~s${}^{-1}$. Our lack of a lens model does not allow us to construct a dynamical model of MRG-P0918, but we can place a lower limit on its rotational support and classify this galaxy as a ``fast rotator.''

\subsection{Dynamical Masses}

We find high dynamical masses for all three galaxies, spanning the range $\log M_{\rm dyn} = 11.26$-11.64. The stellar-to-dynamical mass ratios are $\log M_* / M_{\rm dyn} = 0.05 \pm 0.20$ for MRG-M0138, $0.24 \pm 0.21$ for MRG-M0150, and $-0.42 \pm 0.13$ for MRG-M2129. (These include the magnification uncertainty.) This range is similar to that obtained from unresolved kinematics of $z\sim 2$ quiescent galaxies \citep{Belli17}.

MRG-M0138 and MRG-M0150 are consistent with equality between the dynamical and stellar masses, assuming a \citet{Chabrier03} initial mass function (IMF). The stellar mass of MRG-M2129 is the best constrained and is significantly lighter than the dynamical mass, leaving room for additional dark matter, gas, or a heavier IMF. Given the uncertainties, both MRG-M0138 and MRG-M2129 could also support a Salpeter IMF, which would be in tension with the dynamical mass of MRG-M0150. Future observations of high-$z$ quiescent galaxies might be able to constrain the normalization of the IMF close to the epoch of the star formation, when the stellar population is less polluted by mergers, but such comparisons will have to wait for larger samples with well-resolved kinematics.

\section{Discussion}

All four of the massive quiescent galaxies at $z=1.95$-2.64 in our study are classified as ``fast rotators'' \citep{Emsellem11}. The three systems for which the lensing constraints permit a source reconstruction appear to have similar kinematics: all are rotationally supported galaxies with $V_{\rm max} = 290$-352~km~s${}^{-1}$ and $(V/\sigma)_{R_e} = 1.6$-2.3. For MRG-M0150, this result depends on model-dependent inclination estimates (Section~\ref{sec:results}), while MRG-M0138 and MRG-M2129 are unambiguous. Interestingly MRG-M0150 would appear to be the least disk-like galaxy judging from its low projected ellipticity and high S\'{e}rsic index ($n=3.5$). Only with stellar kinematic data can we show that it is likely to be a rapidly rotating galaxy viewed at low inclination.

As we will discuss, the degree of rotational support that we find in the lensed sample is much higher than is seen in local early-type galaxies that are sufficiently massive to represent their likely descendants. These lensed galaxies  must evolve substantially in their structure and kinematics after quenching. After discussing the evolution of these galaxies into local early-type systems (Section~\ref{sec:disc_local}), we will consider the implications for quenching and the star-forming progenitors of massive quiescent galaxies at $z \sim 2$ (Section~\ref{sec:disc_highz}). When considering these implications, an important caveat to bear in mind is that our study is based on a small sample in which high-ellipticity galaxies are over-represented (Paper~I). While the similar dynamical structures of the galaxies in our sample suggest that they represent a typical case, a larger sample is ultimately needed to assess the prevalence of disk-dominated dynamics in high-$z$ quiescent galaxies.

\subsection{Evolution into $z \sim 0$ Early-type Galaxies}
\label{sec:disc_local}

Figure~\ref{fig:lambda} compares the angular momentum parameter $\lambda_{R_e}$ of the three galaxies in our sample for which we have a lens model to a sample of local early-type galaxies drawn from the \ATD~\citep{Cappellari11,Emsellem11} and MASSIVE \citep{Ma14,Veale17} surveys. Galaxies with $\log M/\msol > 11.2$ are plotted; all of the high-$z$ galaxies have $M_{\rm dyn}$ in this range.\footnote{We use dynamical masses for the lensed galaxies and \ATD. Since these are not available for the MASSIVE sample, we instead use the MASSIVE galaxies' stellar masses, estimated from the $K$-band luminosity following Equation 1 of \citet{Veale17a}. This is appropriate since the two masses agree on average for high-mass galaxies in \ATD.}

It is immediately apparent that the high-$z$ galaxies have much more rotation than typical early-type galaxies in the local universe that are comparably massive. The rarity of local analogs of the galaxies in our sample allows us to make a strong inference about their future evolution. Among galaxies in the \ATD~and MASSIVE samples that are at least as massive as MRG-M0150, 16\% have a higher $\lambda_{R_e}$.\footnote{When quoting properties of the combined \ATD~and MASSIVE samples, we weight the MASSIVE galaxies by a factor of 0.08 relative to \ATD~to account for the different volumes and completeness of the surveys. This is necessary to ensure that galaxies of different masses are represented in the correct proportion. The weight is based on the relative number of MASSIVE galaxies (75 with $\lambda_{R_e}$ from \citealt{Veale17}) to \ATD~galaxies (6) with $K$-band fluxes above the MASSIVE limit \citep{Ma14}.} This implies that MRG-M0150 is atypical among possible descendants in this parameter, but not extremely so. On the other hand, among galaxies in the local sample that are at least as massive as MRG-M0138 (79 galaxies) or MRG-M2129 (92 galaxies), none has a higher $\lambda_{R_e}$ \emph{or} a higher projected ellipticity ($e = 0.74$ and 0.71 for MRG-M0138 and MRG-M2129, respectively; see Paper~I). Although these galaxies are both viewed nearly edge on, this rarity is not primarily an inclination effect: the model grid in Figure~\ref{fig:lambda} implies that they are also intrinsically flatter than all but one of the local sample.

The comoving number density of $M_* = 10^{11-11.5} \msol$ quiescent galaxies at $z=2$ is $\simeq 10$-20\% of the value at $z \sim 0$ \citep{Moustakas13,Tomczak14}. Therefore, unless MRG-M0138 and MRG-M2129 were very atypical of $z \sim 2$ quiescent galaxies, we would expect to find a significant number of local analogs if $\lambda_{R_e}$ did not decline after quenching. We conclude that a significant fraction of massive quiescent galaxies at $z \sim 2$ (the majority of our small sample) must decrease their specific angular momentum and become rounder.

By how much has $\lambda_{R_e}$ declined in such galaxies? Since they likely follow a wide range of evolutionary paths, we can only estimate a rough number by assuming that our sample is representative both of $z \sim 2$ quiescent galaxies and of the progenitors of some $z \sim 0$ comparison population. Simulations and empirical arguments based on number densities suggest that galaxies in the mass range of our sample have grown in mass since $z \sim 2$ by 0.3~dex, on average, with a wide dispersion \citep{vanDokkum10,Muzzin13,Wellons15}. We therefore selected a comparison sample of local galaxies with masses that are 0.3~dex higher than each lensed quiescent galaxy. (In practice, in order to have adequate statistics, we chose galaxies within $\pm 0.15$~dex of that mass.)  For MRG-M0138, MRG-M0150, and MRG-M2129, the median $\lambda_{R_e}$ of these candidate descendants is lower by a factor of 14, 6, and 10, respectively. This result is not very sensitive to the definition of the local comparison sample; for example, we find lower $\lambda_{R_e}$ by factors of 10, 5, and 8, respectively, if we simply select local galaxies that are at least as massive as each high-$z$ galaxy. We therefore estimate that massive quiescent galaxies at $z \sim 2$ have declined in $\lambda_{R_e}$ by a typical factor of 5-10.

Mergers are the likely mechanism to accomplish such a large reduction in spin. In recent years the emergence of local early-type galaxies and their dynamical properties has been investigated by several groups using cosmological simulations \citep{Naab14,Penoyre17,Lagos18b,Lagos18a}. Although these works disagree on some aspects (e.g., the relative important of mass ratio versus gas fraction in setting the remnant spin), many of the broad trends are common. Slow rotators were born as fast rotators, and few slow rotators are expected to be present at $z \gtrsim 1$. Mergers are the primary cause of the loss of angular momentum over time that today's slow rotators experienced. Major mergers tend to reduce angular momentum suddenly, although rare configurations can instead ``spin up'' the remnant. However, a gradual transformation by minor mergers is also possible, with the total merged mass emerging as the most important parameter in some studies. \citet{Bournaud07} presented simulations of a disk galaxy undergoing a series of minor mergers that double its mass and reduce $V/\sigma$ from 3 to $\lesssim 0.2$. Similarly \citet{Naab14} discuss a subset of galaxies in cosmological simulations whose assembly histories are dominated by gas-poor minor mergers and whose angular momentum declined by a factor of $\simeq 10$ since $z=2$ (their Figure~3, Class F), consistent with our estimates. Therefore major mergers are probably important but are not required to produce the observed spin down. Finally, although mergers are a key factor shaping the spin history of most systems, \citet{Lagos18b} found that 30\% of $z=0$ slow rotators in the Illustris simulation have had no significant mergers. They suggested that such galaxies instead inherited their low spin from their dark matter halos.

A rare subset of galaxies may have remained virtually untouched since $z \sim 2$. Searches have uncovered a few massive galaxies with old stellar populations and compact sizes (similar to high-$z$ quiescent galaxies) in the local universe, including NGC1277 \citep{vandenBosch12,Trujillo14}, PGC032873, and Mrk1216 \citep{Ferre-Mateu17}.\footnote{The only local galaxy in Figure~\ref{fig:lambda} that falls within the error bars of one of the lensed galaxies is NGC1167, a very early-type spiral with similar kinematics to MRG-M0150 but with an extended stellar disk having $R_e \sim 13$~kpc \citep{Marchuk17}. NGC1167 is therefore not a high-$z$ analog.} Interestingly, these galaxies have projected rotation speeds of $V \simeq 200$-300~km~s${}^{-1}$ and projected $V/\sigma \simeq 1$-2. The similarity of their kinematics to the high-$z$ galaxies in our sample supports the idea that these are indeed ``relic'' galaxies. Such ``relic'' galaxies appear to be rare, as expected from cosmological merger histories \citep{Quilis13}. 

\citet{Bezanson18} recently measured the rotational support of 104 quiescent galaxies at $z \sim 0.8$. By comparing to a homogeneously analyzed $z \sim 0$ sample, they infer a decrease in rotational support since $z \sim 0.8$ by a factor of $\sim 2$. Given that we estimated a factor of 5-10 decline since $z \sim 2$ for massive galaxies ($M_{\rm dyn, z\sim 2} \gtrsim 10^{11.3} \msol$), this suggests that much of the spin down might have occurred at $z > 1$. That would be qualitatively consistent with the evolution of the mass--size relation, which is also thought to be driven by mergers and also evolves by a similar factor over $z=1$-2 as over $z=0$-1 (e.g., \citealt{Newman12}) . However, it is not consistent with the predictions of cosmological simulations, in which angular momentum loss mainly occurs at $z < 1$ \citep{Penoyre17,Lagos18b}. A robust test of that prediction will require larger samples of $z > 1$ stellar kinematics measured with the \emph{James Webb Space Telescope} or 30~m-class ground-based telescopes.

Finally, the ``spin down'' of quiescent galaxies may provide a new way to distinguish post-quenching evolution from progenitor effects. Since star-forming galaxies are continually growing in size, as they quench and join the quiescent population they will increase the mean size of quiescent systems \citep[e.g.,][]{Carollo13}. Disentangling this growth---which occurs in star-forming galaxies---from size growth in quiescent systems is challenging \citep[e.g.,][]{Belli15}. The situation is different for $V/\sigma$, because it increases over time in star-forming galaxies \citep[e.g.,][]{Wisnioski15,Simons17}, whereas this paper and \citet{Bezanson18} show that $V/\sigma$ declines in quiescent galaxies. Processes that are coincident with or follow quenching are therefore needed to decrease $V/\sigma$ in the quiescent population. The rarity of local analogs of our sample argues that post-quenching ``spin down'' must occur. At the high masses sampled by our survey, the importance of mergers is fairly uncontroversial, but as future observations probe the kinematics of less massive quiescent galaxies at high redshifts, they will provide an interesting new constraint on evolutionary models. 

\subsection{Relation to Star-forming Progenitors}
\label{sec:disc_highz}

Our analysis implies that rotational support generally declines in massive galaxies after quenching. It is interesting to consider whether there is also a more sudden decline in rotational support associated with quenching. This question is harder to address. The kinematics of similarly massive star-forming galaxies that are coeval with our sample have been measured. \citet{Tadaki17a,Tadaki17b} resolved the H$\alpha$ and CO kinematics of 11 extended star-forming galaxies with masses $M_* \gtrsim 10^{11} \msol$. They found $V/\sigma$ spans the range 3.5-7.1, significantly higher than the values of $\simeq2$ we see in quiescent galaxies. Compact star-forming galaxies have been proposed as transitional objects that are the immediate progenitors of compact quiescent galaxies \citep{Barro13}. \citet{vanDokkum15} inferred such galaxies have rapidly rotating ionized gas disks based on marginally resolved H$\alpha$ kinematics. Recently \citet{Barro17} measured the CO kinematics of such a galaxy and found $V/\sigma \simeq 2.5$. This is consistent with our quiescent sample and supports the idea that ``blue nuggets'' are immediate progenitors of some compact quiescent galaxies, assuming that the kinematics of the stars and CO are not too dissimilar.

Assessing a decline in $V/\sigma$ associated with quenching requires comparing not to coeval systems, but to massive star-forming galaxies at $z=3$-4, the quenching epoch of the galaxies in our sample. Surveys of ionized gas kinematics in this redshift range do not sample galaxies that are sufficiently massive to be progenitors of galaxies in our sample \citep{Gnerucci11,Turner17}. CO observations of a few massive starburst galaxies at $z\approx4$-5, which are probably progenitors of some massive quiescent galaxies \citep[e.g.,][]{Toft14}, have shown a range of kinematic properties with both less and more rotational support than our sample \citep{Hodge12,Riechers14,Oteo16}.

Given these few constraints, the lower $V/\sigma$ seen in our quiescent galaxy sample compared to coeval star-forming galaxies could result from two effects. First, $V/\sigma$ might decline when star formation is quenched, but our analysis in Section~\ref{sec:disc_local} indicates that it does not generally reach the lower levels seen in local early-type galaxies, which requires mergers after quenching. Second, the lower $V/\sigma$ of $z\sim2$ quiescent galaxies might be inherited from their star-forming progenitors at $z \sim 3$-4, with less or even no change associated with quenching.

Theoretically it has been shown that gas-rich major mergers can produce remnants with disks \citep[e.g.,][]{Robertson06,Governato09,Wuyts10,Sparre17}. The persistence of disk-dominated kinematics therefore does not require a particularly ``gentle'' quenching process. The \citet{Wuyts10} simulations of binary wet mergers are particularly relevant, since they compared the simulated remnants to the observed properties of $z \sim 2$ quiescent galaxies. Provided that the gas fraction at coalescence is high ($\gtrsim 40\%$), which is expected to occur more frequently at higher redshifts, the remnants can be quiescent galaxies with sizes as compact as observed. Wuyts et al.~also made a prediction that these galaxies are much more rotationally supported than local early-type galaxies. Their simulated remnants lie near or above the isotropic rotator line in the $V/\sigma$ versus ellipticity plot (their Figure 5), broadly consistent with the measurements in this paper. The sizes and kinematics of our sample may therefore be compatible with remnants of gas-rich major mergers.

The very flattened shapes of MRG-M0138 and MRG-M2129, however, present more of a puzzle, since the Wuyts et al.~simulations do not produce such thin remnants. Furthermore, the simulated remnants have cuspy light profiles that are unlike the exponential disk observed in MRG-M2129. This galaxy is particularly interesting, since it demonstrates that it is possible for star formation to quench at $z \sim 3$ without the formation of a significant bulge. It lacks a central enhancement of the stellar mass density (Paper~I, Figure~7) that one might expect in major mergers or ``compaction'' scenarios where a central starburst precedes quenching \citep{Dekel14,Zolotov15}. \citet{Toft17} suggested that MRG-M2129 was instead quenched as gas flowing onto the (presumably massive) halo of MRG-M2129 was shock heated and prevented from accreting onto the galaxy. In this scenario the recently quenched object could retain the structure and kinematics of its star-forming progenitor. Considering their range of bulge properties, it seems likely that massive quiescent galaxies at $z \approx 2$ were formed through multiple channels. 

\section{Summary}

We presented the resolved stellar kinematics of a sample of four lensed quiescent galaxies at $z=1.95$-2.64. All four are classified as ``fast rotators.'' For the three galaxies with lens models that enable a source plane reconstruction, we constructed JAM dynamical models. Their dynamical masses $M_{\rm dyn} \gtrsim 2 \times 10^{11} \msol$ support the high stellar masses reported in Paper~I. Two of these three galaxies are  highly inclined, and we argued that the third (MRG-M0150) is likely seen at a low inclination of $i \approx 30^{\circ}$. With that assumption, all three systems are intrinsically thin ($e_{\rm intr} \approx 0.75-0.85$), rapidly rotating ($V_{\rm max} = 290$-352~km~s${}^{-1}$) galaxies that are primarily rotationally supported with $(V/\sigma)_{R_e}=1.6$-2.3. These $V/\sigma$ values are smaller than massive, coeval star-forming galaxies. This could reflect a decline in rotational support associated with quenching, or alternatively the lower $V/\sigma$ might be inherited from the $z=3$-4 star-forming progenitors.

The galaxies in our sample show much more rotation than typical local quiescent galaxies that have masses consistent with being their descendants. In particular, their angular momentum parameter $\lambda_{R_e}$ is typically 5-10 times higher than a local comparison sample selected from the \ATD~and MASSIVE surveys. For MRG-M0138 and MRG-M2129, both $\lambda_{R_e}$ and the ellipticity are higher than any galaxy in the local comparison sample with an equal or higher mass. If our small sample is representative, these observations show that while rotational support might be eroded in massive galaxies when star formation is quenched, it does not generally reach the lower values characteristic of massive early-type galaxies in the local universe. That transformation occurs after quenching, most likely proceeding via a series of mergers that grow galaxies in mass and size while reducing their specific angular momentum.

\acknowledgments
We thank the anonymous referee for a close reading and helpful suggestions that improved this paper. RSE acknowledges financial support from European Research Council Advanced Grant FP7/669253. Support for programs GO-14496 and GO-14205 was provided by NASA through grants from the Space Telescope Science Institute, which is operated by the Associations of Universities for Research in Astronomy, Incorporated, under NASA contract NAS5-26555. This paper includes data gathered with the 6.5 meter Magellan Telescopes located at Las Campanas Observatory, Chile. Some of the data presented herein were obtained at the W. M. Keck Observatory, which is operated as a scientific partnership among the California Institute of Technology, the University of California and the National Aeronautics and Space Administration. The Observatory was made possible by the generous financial support of the W. M. Keck Foundation. The authors wish to recognize and acknowledge the very significant cultural role and reverence that the summit of Maunakea has always had within the indigenous Hawaiian community.  We are most fortunate to have the opportunity to conduct observations from this mountain.

\bibliographystyle{apj}
\bibliography{lensed_nuggets_II}

\begin{thebibliography}{79}
\expandafter\ifx\csname natexlab\endcsname\relax\def\natexlab#1{#1}\fi

\bibitem[{{Barro} {et~al.}(2013){Barro}, {Faber}, {P{\'e}rez-Gonz{\'a}lez},
  {Koo}, {Williams}, {Kocevski}, {Trump}, {Mozena}, {McGrath}, {van der Wel},
  {Wuyts}, {Bell}, {Croton}, {Ceverino}, {Dekel}, {Ashby}, {Cheung},
  {Ferguson}, {Fontana}, {Fang}, {Giavalisco}, {Grogin}, {Guo}, {Hathi},
  {Hopkins}, {Huang}, {Koekemoer}, {Kartaltepe}, {Lee}, {Newman}, {Porter},
  {Primack}, {Ryan}, {Rosario}, {Somerville}, {Salvato}, \& {Hsu}}]{Barro13}
{Barro}, G., {Faber}, S.~M., {P{\'e}rez-Gonz{\'a}lez}, P.~G., {et~al.} 2013,
  \apj, 765, 104

\bibitem[{{Barro} {et~al.}(2017){Barro}, {Kriek}, {P{\'e}rez-Gonz{\'a}lez},
  {Diaz-Santos}, {Price}, {Rujopakarn}, {Pandya}, {Koo}, {Faber}, {Dekel},
  {Primack}, \& {Kocevski}}]{Barro17}
{Barro}, G., {Kriek}, M., {P{\'e}rez-Gonz{\'a}lez}, P.~G., {et~al.} 2017,
  \apjl, 851, L40

\bibitem[{{Belli} {et~al.}(2015){Belli}, {Newman}, \& {Ellis}}]{Belli15}
{Belli}, S., {Newman}, A.~B., \& {Ellis}, R.~S. 2015, \apj, 799, 206

\bibitem[{{Belli} {et~al.}(2017){Belli}, {Newman}, \& {Ellis}}]{Belli17}
---. 2017, \apj, 834, 18

\bibitem[{{Bezanson} {et~al.}(2009){Bezanson}, {van Dokkum}, {Tal},
  {Marchesini}, {Kriek}, {Franx}, \& {Coppi}}]{Bezanson09}
{Bezanson}, R., {van Dokkum}, P.~G., {Tal}, T., {et~al.} 2009, \apj, 697, 1290

\bibitem[{{Bezanson} {et~al.}(2018){Bezanson}, {van der Wel}, {Pacifici},
  {Noeske}, {Bari{\v s}i{\'c}}, {Bell}, {Brammer}, {Calhau}, {Chauke}, {van
  Dokkum}, {Franx}, {Gallazzi}, {van Houdt}, {Labb{\'e}}, {Maseda},
  {Mu{\~n}os-Mateos}, {Muzzin}, {van de Sande}, {Sobral}, {Straatman}, \&
  {Wu}}]{Bezanson18}
{Bezanson}, R., {van der Wel}, A., {Pacifici}, C., {et~al.} 2018,
  arXiv:1804.02402

\bibitem[{{Bournaud} {et~al.}(2007){Bournaud}, {Jog}, \& {Combes}}]{Bournaud07}
{Bournaud}, F., {Jog}, C.~J., \& {Combes}, F. 2007, \aap, 476, 1179

\bibitem[{{Bruce} {et~al.}(2012){Bruce}, {Dunlop}, {Cirasuolo}, {McLure},
  {Targett}, {Bell}, {Croton}, {Dekel}, {Faber}, {Ferguson}, {Grogin},
  {Kocevski}, {Koekemoer}, {Koo}, {Lai}, {Lotz}, {McGrath}, {Newman}, \& {van
  der Wel}}]{Bruce12}
{Bruce}, V.~A., {Dunlop}, J.~S., {Cirasuolo}, M., {et~al.} 2012, \mnras, 427,
  1666

\bibitem[{{Bruce} {et~al.}(2014){Bruce}, {Dunlop}, {McLure}, {Cirasuolo},
  {Buitrago}, {Bowler}, {Targett}, {Bell}, {McIntosh}, {Dekel}, {Faber},
  {Ferguson}, {Grogin}, {Hartley}, {Kocevski}, {Koekemoer}, {Koo}, \&
  {McGrath}}]{Bruce14}
{Bruce}, V.~A., {Dunlop}, J.~S., {McLure}, R.~J., {et~al.} 2014, \mnras, 444,
  1001

\bibitem[{{Bruzual} \& {Charlot}(2003)}]{BC03}
{Bruzual}, G., \& {Charlot}, S. 2003, \mnras, 344, 1000

\bibitem[{{Buitrago} {et~al.}(2013){Buitrago}, {Trujillo}, {Conselice}, \&
  {H{\"a}u{\ss}ler}}]{Buitrago13}
{Buitrago}, F., {Trujillo}, I., {Conselice}, C.~J., \& {H{\"a}u{\ss}ler}, B.
  2013, \mnras, 428, 1460

\bibitem[{{Cappellari}(2008)}]{Cappellari08}
{Cappellari}, M. 2008, \mnras, 390, 71

\bibitem[{{Cappellari}(2016)}]{Cappellari16}
---. 2016, \araa, 54, 597

\bibitem[{{Cappellari} \& {Emsellem}(2004)}]{Cappellari04}
{Cappellari}, M., \& {Emsellem}, E. 2004, \pasp, 116, 138

\bibitem[{{Cappellari} {et~al.}(2007){Cappellari}, {Emsellem}, {Bacon},
  {Bureau}, {Davies}, {de Zeeuw}, {Falc{\'o}n-Barroso}, {Krajnovi{\'c}},
  {Kuntschner}, {McDermid}, {Peletier}, {Sarzi}, {van den Bosch}, \& {van de
  Ven}}]{Cappellari07}
{Cappellari}, M., {Emsellem}, E., {Bacon}, R., {et~al.} 2007, \mnras, 379, 418

\bibitem[{{Cappellari} {et~al.}(2011){Cappellari}, {Emsellem}, {Krajnovi{\'c}},
  {McDermid}, {Scott}, {Verdoes Kleijn}, {Young}, {Alatalo}, {Bacon}, {Blitz},
  {Bois}, {Bournaud}, {Bureau}, {Davies}, {Davis}, {de Zeeuw}, {Duc},
  {Khochfar}, {Kuntschner}, {Lablanche}, {Morganti}, {Naab}, {Oosterloo},
  {Sarzi}, {Serra}, \& {Weijmans}}]{Cappellari11}
{Cappellari}, M., {Emsellem}, E., {Krajnovi{\'c}}, D., {et~al.} 2011, \mnras,
  413, 813

\bibitem[{{Cappellari} {et~al.}(2013){Cappellari}, {Scott}, {Alatalo}, {Blitz},
  {Bois}, {Bournaud}, {Bureau}, {Crocker}, {Davies}, {Davis}, {de Zeeuw},
  {Duc}, {Emsellem}, {Khochfar}, {Krajnovi{\'c}}, {Kuntschner}, {McDermid},
  {Morganti}, {Naab}, {Oosterloo}, {Sarzi}, {Serra}, {Weijmans}, \&
  {Young}}]{Cappellari13}
{Cappellari}, M., {Scott}, N., {Alatalo}, K., {et~al.} 2013, \mnras, 432, 1709

\bibitem[{{Carollo} {et~al.}(2013){Carollo}, {Bschorr}, {Renzini}, {Lilly},
  {Capak}, {Cibinel}, {Ilbert}, {Onodera}, {Scoville}, {Cameron}, {Mobasher},
  {Sanders}, \& {Taniguchi}}]{Carollo13}
{Carollo}, C.~M., {Bschorr}, T.~J., {Renzini}, A., {et~al.} 2013, \apj, 773,
  112

\bibitem[{{Chabrier}(2003)}]{Chabrier03}
{Chabrier}, G. 2003, \pasp, 115, 763

\bibitem[{{Chang} {et~al.}(2013{\natexlab{a}}){Chang}, {van der Wel}, {Rix},
  {Wuyts}, {Zibetti}, {Ramkumar}, \& {Holden}}]{Chang13a}
{Chang}, Y.-Y., {van der Wel}, A., {Rix}, H.-W., {et~al.} 2013{\natexlab{a}},
  \apj, 762, 83

\bibitem[{{Chang} {et~al.}(2013{\natexlab{b}}){Chang}, {van der Wel}, {Rix},
  {Holden}, {Bell}, {McGrath}, {Wuyts}, {H{\"a}ussler}, {Barden}, {Faber},
  {Mozena}, {Ferguson}, {Guo}, {Galametz}, {Grogin}, {Kocevski}, {Koekemoer},
  {Dekel}, {Huang}, {Hathi}, \& {Donley}}]{Chang13b}
---. 2013{\natexlab{b}}, \apj, 773, 149

\bibitem[{{Conroy} \& {Gunn}(2010)}]{Conroy10}
{Conroy}, C., \& {Gunn}, J.~E. 2010, \apj, 712, 833

\bibitem[{{Conroy} {et~al.}(2009){Conroy}, {Gunn}, \& {White}}]{Conroy09}
{Conroy}, C., {Gunn}, J.~E., \& {White}, M. 2009, \apj, 699, 486

\bibitem[{{Dekel} \& {Burkert}(2014)}]{Dekel14}
{Dekel}, A., \& {Burkert}, A. 2014, \mnras, 438, 1870

\bibitem[{{Emsellem} {et~al.}(2007){Emsellem}, {Cappellari}, {Krajnovi{\'c}},
  {van de Ven}, {Bacon}, {Bureau}, {Davies}, {de Zeeuw}, {Falc{\'o}n-Barroso},
  {Kuntschner}, {McDermid}, {Peletier}, \& {Sarzi}}]{Emsellem07}
{Emsellem}, E., {Cappellari}, M., {Krajnovi{\'c}}, D., {et~al.} 2007, \mnras,
  401

\bibitem[{{Emsellem} {et~al.}(2011){Emsellem}, {Cappellari}, {Krajnovi{\'c}},
  {Alatalo}, {Blitz}, {Bois}, {Bournaud}, {Bureau}, {Davies}, {Davis}, {de
  Zeeuw}, {Khochfar}, {Kuntschner}, {Lablanche}, {McDermid}, {Morganti},
  {Naab}, {Oosterloo}, {Sarzi}, {Scott}, {Serra}, {van de Ven}, {Weijmans}, \&
  {Young}}]{Emsellem11}
---. 2011, \mnras, 414, 888

\bibitem[{{Ferr{\'e}-Mateu} {et~al.}(2017){Ferr{\'e}-Mateu}, {Trujillo},
  {Mart{\'{\i}}n-Navarro}, {Vazdekis}, {Mezcua}, {Balcells}, \&
  {Dom{\'{\i}}nguez}}]{Ferre-Mateu17}
{Ferr{\'e}-Mateu}, A., {Trujillo}, I., {Mart{\'{\i}}n-Navarro}, I., {et~al.}
  2017, \mnras, 467, 1929

\bibitem[{{Foreman-Mackey}(2017)}]{cornerpy}
{Foreman-Mackey}, D. 2017, {corner.py: Corner plots}, Astrophysics Source Code
  Library

\bibitem[{{Geier} {et~al.}(2013){Geier}, {Richard}, {Man}, {Kr{\"u}hler},
  {Toft}, {Marchesini}, \& {Fynbo}}]{Geier13}
{Geier}, S., {Richard}, J., {Man}, A.~W.~S., {et~al.} 2013, \apj, 777, 87

\bibitem[{{Gnerucci} {et~al.}(2011){Gnerucci}, {Marconi}, {Cresci}, {Maiolino},
  {Mannucci}, {Calura}, {Cimatti}, {Cocchia}, {Grazian}, {Matteucci}, {Nagao},
  {Pozzetti}, \& {Troncoso}}]{Gnerucci11}
{Gnerucci}, A., {Marconi}, A., {Cresci}, G., {et~al.} 2011, \aap, 528, A88

\bibitem[{{Governato} {et~al.}(2009){Governato}, {Brook}, {Brooks}, {Mayer},
  {Willman}, {Jonsson}, {Stilp}, {Pope}, {Christensen}, {Wadsley}, \&
  {Quinn}}]{Governato09}
{Governato}, F., {Brook}, C.~B., {Brooks}, A.~M., {et~al.} 2009, \mnras, 398,
  312

\bibitem[{{Hodge} {et~al.}(2012){Hodge}, {Carilli}, {Walter}, {de Blok},
  {Riechers}, {Daddi}, \& {Lentati}}]{Hodge12}
{Hodge}, J.~A., {Carilli}, C.~L., {Walter}, F., {et~al.} 2012, \apj, 760, 11

\bibitem[{{Hopkins} {et~al.}(2010){Hopkins}, {Bundy}, {Hernquist}, {Wuyts}, \&
  {Cox}}]{Hopkins10}
{Hopkins}, P.~F., {Bundy}, K., {Hernquist}, L., {Wuyts}, S., \& {Cox}, T.~J.
  2010, \mnras, 401, 1099

\bibitem[{{Lagos} {et~al.}(2018{\natexlab{a}}){Lagos}, {Schaye}, {Bah{\'e}},
  {Van de Sande}, {Kay}, {Barnes}, {Davis}, \& {Dalla Vecchia}}]{Lagos18b}
{Lagos}, C.~d.~P., {Schaye}, J., {Bah{\'e}}, Y., {et~al.} 2018{\natexlab{a}},
  \mnras, 476, 4327

\bibitem[{{Lagos} {et~al.}(2018{\natexlab{b}}){Lagos}, {Stevens}, {Bower},
  {Davis}, {Contreras}, {Padilla}, {Obreschkow}, {Croton}, {Trayford},
  {Welker}, \& {Theuns}}]{Lagos18a}
{Lagos}, C.~d.~P., {Stevens}, A.~R.~H., {Bower}, R.~G., {et~al.}
  2018{\natexlab{b}}, \mnras, 473, 4956

\bibitem[{{Ma} {et~al.}(2014){Ma}, {Greene}, {McConnell}, {Janish},
  {Blakeslee}, {Thomas}, \& {Murphy}}]{Ma14}
{Ma}, C.-P., {Greene}, J.~E., {McConnell}, N., {et~al.} 2014, \apj, 795, 158

\bibitem[{{Marchuk} \& {Sotnikova}(2017)}]{Marchuk17}
{Marchuk}, A.~A., \& {Sotnikova}, N.~Y. 2017, \mnras, 465, 4956

\bibitem[{{McGrath} {et~al.}(2008){McGrath}, {Stockton}, {Canalizo}, {Iye}, \&
  {Maihara}}]{McGrath08}
{McGrath}, E.~J., {Stockton}, A., {Canalizo}, G., {Iye}, M., \& {Maihara}, T.
  2008, \apj, 682, 303

\bibitem[{{McLean} {et~al.}(2012){McLean}, {Steidel}, {Epps}, {Konidaris},
  {Matthews}, {Adkins}, {Aliado}, {Brims}, {Canfield}, {Cromer}, {Fucik},
  {Kulas}, {Mace}, {Magnone}, {Rodriguez}, {Rudie}, {Trainor}, {Wang}, {Weber},
  \& {Weiss}}]{McLean12}
{McLean}, I.~S., {Steidel}, C.~C., {Epps}, H.~W., {et~al.} 2012, in Society of
  Photo-Optical Instrumentation Engineers (SPIE) Conference Series, Vol. 8446,
  Society of Photo-Optical Instrumentation Engineers (SPIE) Conference Series,
  0

\bibitem[{{McLure} {et~al.}(2013){McLure}, {Pearce}, {Dunlop}, {Cirasuolo},
  {Curtis-Lake}, {Bruce}, {Caputi}, {Almaini}, {Bonfield}, {Bradshaw},
  {Buitrago}, {Chuter}, {Foucaud}, {Hartley}, \& {Jarvis}}]{McLure13}
{McLure}, R.~J., {Pearce}, H.~J., {Dunlop}, J.~S., {et~al.} 2013, \mnras, 428,
  1088

\bibitem[{{Moustakas} {et~al.}(2013){Moustakas}, {Coil}, {Aird}, {Blanton},
  {Cool}, {Eisenstein}, {Mendez}, {Wong}, {Zhu}, \& {Arnouts}}]{Moustakas13}
{Moustakas}, J., {Coil}, A.~L., {Aird}, J., {et~al.} 2013, \apj, 767, 50

\bibitem[{{Muzzin} {et~al.}(2013){Muzzin}, {Marchesini}, {Stefanon}, {Franx},
  {McCracken}, {Milvang-Jensen}, {Dunlop}, {Fynbo}, {Brammer}, {Labb{\'e}}, \&
  {van Dokkum}}]{Muzzin13}
{Muzzin}, A., {Marchesini}, D., {Stefanon}, M., {et~al.} 2013, \apj, 777, 18

\bibitem[{{Naab} {et~al.}(2009){Naab}, {Johansson}, \& {Ostriker}}]{Naab09}
{Naab}, T., {Johansson}, P.~H., \& {Ostriker}, J.~P. 2009, \apjl, 699, L178

\bibitem[{{Naab} {et~al.}(2014){Naab}, {Oser}, {Emsellem}, {Cappellari},
  {Krajnovi{\'c}}, {McDermid}, {Alatalo}, {Bayet}, {Blitz}, {Bois}, {Bournaud},
  {Bureau}, {Crocker}, {Davies}, {Davis}, {de Zeeuw}, {Duc}, {Hirschmann},
  {Johansson}, {Khochfar}, {Kuntschner}, {Morganti}, {Oosterloo}, {Sarzi},
  {Scott}, {Serra}, {van de Ven}, {Weijmans}, \& {Young}}]{Naab14}
{Naab}, T., {Oser}, L., {Emsellem}, E., {et~al.} 2014, \mnras, 444, 3357

\bibitem[{{Newman} {et~al.}(2015){Newman}, {Belli}, \& {Ellis}}]{Newman15}
{Newman}, A.~B., {Belli}, S., \& {Ellis}, R.~S. 2015, \apjl, 813, L7

\bibitem[{{Newman} {et~al.}(2014){Newman}, {Ellis}, {Andreon}, {Treu},
  {Raichoor}, \& {Trinchieri}}]{Newman14}
{Newman}, A.~B., {Ellis}, R.~S., {Andreon}, S., {et~al.} 2014, \apj, 788, 51

\bibitem[{{Newman} {et~al.}(2012){Newman}, {Ellis}, {Bundy}, \&
  {Treu}}]{Newman12}
{Newman}, A.~B., {Ellis}, R.~S., {Bundy}, K., \& {Treu}, T. 2012, \apj, 746,
  162

\bibitem[{{Oser} {et~al.}(2010){Oser}, {Ostriker}, {Naab}, {Johansson}, \&
  {Burkert}}]{Oser10}
{Oser}, L., {Ostriker}, J.~P., {Naab}, T., {Johansson}, P.~H., \& {Burkert}, A.
  2010, \apj, 725, 2312

\bibitem[{{Oteo} {et~al.}(2016){Oteo}, {Ivison}, {Dunne}, {Smail}, {Swinbank},
  {Zhang}, {Lewis}, {Maddox}, {Riechers}, {Serjeant}, {Van der Werf}, {Biggs},
  {Bremer}, {Cigan}, {Clements}, {Cooray}, {Dannerbauer}, {Eales}, {Ibar},
  {Messias}, {Micha{\l}owski}, {P{\'e}rez-Fournon}, \& {van Kampen}}]{Oteo16}
{Oteo}, I., {Ivison}, R.~J., {Dunne}, L., {et~al.} 2016, \apj, 827, 34

\bibitem[{{Patel} {et~al.}(2013){Patel}, {van Dokkum}, {Franx}, {Quadri},
  {Muzzin}, {Marchesini}, {Williams}, {Holden}, \& {Stefanon}}]{Patel13}
{Patel}, S.~G., {van Dokkum}, P.~G., {Franx}, M., {et~al.} 2013, \apj, 766, 15

\bibitem[{{Penoyre} {et~al.}(2017){Penoyre}, {Moster}, {Sijacki}, \&
  {Genel}}]{Penoyre17}
{Penoyre}, Z., {Moster}, B.~P., {Sijacki}, D., \& {Genel}, S. 2017, \mnras,
  468, 3883

\bibitem[{{Quilis} \& {Trujillo}(2013)}]{Quilis13}
{Quilis}, V., \& {Trujillo}, I. 2013, \apjl, 773, L8

\bibitem[{{Riechers} {et~al.}(2014){Riechers}, {Carilli}, {Capak}, {Scoville},
  {Smol{\v c}i{\'c}}, {Schinnerer}, {Yun}, {Cox}, {Bertoldi}, {Karim}, \&
  {Yan}}]{Riechers14}
{Riechers}, D.~A., {Carilli}, C.~L., {Capak}, P.~L., {et~al.} 2014, \apj, 796,
  84

\bibitem[{{Robertson} {et~al.}(2006){Robertson}, {Bullock}, {Cox}, {Di Matteo},
  {Hernquist}, {Springel}, \& {Yoshida}}]{Robertson06}
{Robertson}, B., {Bullock}, J.~S., {Cox}, T.~J., {et~al.} 2006, \apj, 645, 986

\bibitem[{{Satoh}(1980)}]{Satoh80}
{Satoh}, C. 1980, \pasj, 32, 41

\bibitem[{{Simcoe} {et~al.}(2013){Simcoe}, {Burgasser}, {Schechter}, {Fishner},
  {Bernstein}, {Bigelow}, {Pipher}, {Forrest}, {McMurtry}, {Smith}, \&
  {Bochanski}}]{Simcoe13}
{Simcoe}, R.~A., {Burgasser}, A.~J., {Schechter}, P.~L., {et~al.} 2013, \pasp,
  125, 270

\bibitem[{{Simons} {et~al.}(2017){Simons}, {Kassin}, {Weiner}, {Faber},
  {Trump}, {Heckman}, {Koo}, {Pacifici}, {Primack}, {Snyder}, \& {de la
  Vega}}]{Simons17}
{Simons}, R.~C., {Kassin}, S.~A., {Weiner}, B.~J., {et~al.} 2017, \apj, 843, 46

\bibitem[{{Sparre} \& {Springel}(2017)}]{Sparre17}
{Sparre}, M., \& {Springel}, V. 2017, \mnras, 470, 3946

\bibitem[{{Stockton} {et~al.}(2008){Stockton}, {McGrath}, {Canalizo}, {Iye}, \&
  {Maihara}}]{Stockton08}
{Stockton}, A., {McGrath}, E., {Canalizo}, G., {Iye}, M., \& {Maihara}, T.
  2008, \apj, 672, 146

\bibitem[{{Tadaki} {et~al.}(2017{\natexlab{a}}){Tadaki}, {Genzel}, {Kodama},
  {Wuyts}, {Wisnioski}, {F{\"o}rster Schreiber}, {Burkert}, {Lang}, {Tacconi},
  {Lutz}, {Belli}, {Davies}, {Hatsukade}, {Hayashi}, {Herrera-Camus},
  {Ikarashi}, {Inoue}, {Kohno}, {Koyama}, {Mendel}, {Nakanishi}, {Shimakawa},
  {Suzuki}, {Tamura}, {Tanaka}, {{\"U}bler}, \& {Wilman}}]{Tadaki17a}
{Tadaki}, K.-i., {Genzel}, R., {Kodama}, T., {et~al.} 2017{\natexlab{a}}, \apj,
  834, 135

\bibitem[{{Tadaki} {et~al.}(2017{\natexlab{b}}){Tadaki}, {Kodama}, {Nelson},
  {Belli}, {F{\"o}rster Schreiber}, {Genzel}, {Hayashi}, {Herrera-Camus},
  {Koyama}, {Lang}, {Lutz}, {Shimakawa}, {Tacconi}, {{\"U}bler}, {Wisnioski},
  {Wuyts}, {Hatsukade}, {Lippa}, {Nakanishi}, {Ikarashi}, {Kohno}, {Suzuki},
  {Tamura}, \& {Tanaka}}]{Tadaki17b}
{Tadaki}, K.-i., {Kodama}, T., {Nelson}, E.~J., {et~al.} 2017{\natexlab{b}},
  \apjl, 841, L25

\bibitem[{{Toft} {et~al.}(2014){Toft}, {Smol{\v c}i{\'c}}, {Magnelli}, {Karim},
  {Zirm}, {Michalowski}, {Capak}, {Sheth}, {Schawinski}, {Krogager}, {Wuyts},
  {Sanders}, {Man}, {Lutz}, {Staguhn}, {Berta}, {Mccracken}, {Krpan}, \&
  {Riechers}}]{Toft14}
{Toft}, S., {Smol{\v c}i{\'c}}, V., {Magnelli}, B., {et~al.} 2014, \apj, 782,
  68

\bibitem[{{Toft} {et~al.}(2017){Toft}, {Zabl}, {Richard}, {Gallazzi},
  {Zibetti}, {Prescott}, {Grillo}, {Man}, {Lee}, {G{\'o}mez-Guijarro},
  {Stockmann}, {Magdis}, \& {Steinhardt}}]{Toft17}
{Toft}, S., {Zabl}, J., {Richard}, J., {et~al.} 2017, \nat, 546, 510

\bibitem[{{Tomczak} {et~al.}(2014){Tomczak}, {Quadri}, {Tran}, {Labb{\'e}},
  {Straatman}, {Papovich}, {Glazebrook}, {Allen}, {Brammer}, {Kacprzak},
  {Kawinwanichakij}, {Kelson}, {McCarthy}, {Mehrtens}, {Monson}, {Persson},
  {Spitler}, {Tilvi}, \& {van Dokkum}}]{Tomczak14}
{Tomczak}, A.~R., {Quadri}, R.~F., {Tran}, K.-V.~H., {et~al.} 2014, \apj, 783,
  85

\bibitem[{{Trujillo} {et~al.}(2014){Trujillo}, {Ferr{\'e}-Mateu}, {Balcells},
  {Vazdekis}, \& {S{\'a}nchez-Bl{\'a}zquez}}]{Trujillo14}
{Trujillo}, I., {Ferr{\'e}-Mateu}, A., {Balcells}, M., {Vazdekis}, A., \&
  {S{\'a}nchez-Bl{\'a}zquez}, P. 2014, \apjl, 780, L20

\bibitem[{{Trujillo} {et~al.}(2006){Trujillo}, {F{\"o}rster Schreiber},
  {Rudnick}, {Barden}, {Franx}, {Rix}, {Caldwell}, {McIntosh}, {Toft},
  {H{\"a}ussler}, {Zirm}, {van Dokkum}, {Labb{\'e}}, {Moorwood},
  {R{\"o}ttgering}, {van der Wel}, {van der Werf}, \& {van
  Starkenburg}}]{Trujillo06}
{Trujillo}, I., {F{\"o}rster Schreiber}, N.~M., {Rudnick}, G., {et~al.} 2006,
  \apj, 650, 18

\bibitem[{{Turner} {et~al.}(2017){Turner}, {Cirasuolo}, {Harrison}, {McLure},
  {Dunlop}, {Swinbank}, {Johnson}, {Sobral}, {Matthee}, \&
  {Sharples}}]{Turner17}
{Turner}, O.~J., {Cirasuolo}, M., {Harrison}, C.~M., {et~al.} 2017, \mnras,
  471, 1280

\bibitem[{{van den Bosch} {et~al.}(2012){van den Bosch}, {Gebhardt},
  {G{\"u}ltekin}, {van de Ven}, {van der Wel}, \& {Walsh}}]{vandenBosch12}
{van den Bosch}, R.~C.~E., {Gebhardt}, K., {G{\"u}ltekin}, K., {et~al.} 2012,
  \nat, 491, 729

\bibitem[{{van der Wel} {et~al.}(2011){van der Wel}, {Rix}, {Wuyts}, {McGrath},
  {Koekemoer}, {Bell}, {Holden}, {Robaina}, \& {McIntosh}}]{vanderWel11}
{van der Wel}, A., {Rix}, H.-W., {Wuyts}, S., {et~al.} 2011, \apj, 730, 38

\bibitem[{{van Dokkum} {et~al.}(2008){van Dokkum}, {Franx}, {Kriek}, {Holden},
  {Illingworth}, {Magee}, {Bouwens}, {Marchesini}, {Quadri}, {Rudnick},
  {Taylor}, \& {Toft}}]{vanDokkum08}
{van Dokkum}, P.~G., {Franx}, M., {Kriek}, M., {et~al.} 2008, \apjl, 677, L5

\bibitem[{{van Dokkum} {et~al.}(2010){van Dokkum}, {Whitaker}, {Brammer},
  {Franx}, {Kriek}, {Labb{\'e}}, {Marchesini}, {Quadri}, {Bezanson},
  {Illingworth}, {Muzzin}, {Rudnick}, {Tal}, \& {Wake}}]{vanDokkum10}
{van Dokkum}, P.~G., {Whitaker}, K.~E., {Brammer}, G., {et~al.} 2010, \apj,
  709, 1018

\bibitem[{{van Dokkum} {et~al.}(2015){van Dokkum}, {Nelson}, {Franx}, {Oesch},
  {Momcheva}, {Brammer}, {F{\"o}rster Schreiber}, {Skelton}, {Whitaker}, {van
  der Wel}, {Bezanson}, {Fumagalli}, {Illingworth}, {Kriek}, {Leja}, \&
  {Wuyts}}]{vanDokkum15}
{van Dokkum}, P.~G., {Nelson}, E.~J., {Franx}, M., {et~al.} 2015, \apj, 813, 23

\bibitem[{{Vazdekis} {et~al.}(2015){Vazdekis}, {Coelho}, {Cassisi},
  {Ricciardelli}, {Falc{\'o}n-Barroso}, {S{\'a}nchez-Bl{\'a}zquez}, {La
  Barbera}, {Beasley}, \& {Pietrinferni}}]{Vazdekis15}
{Vazdekis}, A., {Coelho}, P., {Cassisi}, S., {et~al.} 2015, \mnras, 449, 1177

\bibitem[{{Veale} {et~al.}(2017{\natexlab{a}}){Veale}, {Ma}, {Greene},
  {Thomas}, {Blakeslee}, {McConnell}, {Walsh}, \& {Ito}}]{Veale17}
{Veale}, M., {Ma}, C.-P., {Greene}, J.~E., {et~al.} 2017{\natexlab{a}}, \mnras,
  471, 1428

\bibitem[{{Veale} {et~al.}(2017{\natexlab{b}}){Veale}, {Ma}, {Thomas},
  {Greene}, {McConnell}, {Walsh}, {Ito}, {Blakeslee}, \& {Janish}}]{Veale17a}
{Veale}, M., {Ma}, C.-P., {Thomas}, J., {et~al.} 2017{\natexlab{b}}, \mnras,
  464, 356

\bibitem[{{Wellons} {et~al.}(2015){Wellons}, {Torrey}, {Ma}, {Rodriguez-Gomez},
  {Vogelsberger}, {Kriek}, {van Dokkum}, {Nelson}, {Genel}, {Pillepich},
  {Springel}, {Sijacki}, {Snyder}, {Nelson}, {Sales}, \&
  {Hernquist}}]{Wellons15}
{Wellons}, S., {Torrey}, P., {Ma}, C.-P., {et~al.} 2015, \mnras, 449, 361

\bibitem[{{Wisnioski} {et~al.}(2015){Wisnioski}, {F{\"o}rster Schreiber},
  {Wuyts}, {Wuyts}, {Bandara}, {Wilman}, {Genzel}, {Bender}, {Davies},
  {Fossati}, {Lang}, {Mendel}, {Beifiori}, {Brammer}, {Chan}, {Fabricius},
  {Fudamoto}, {Kulkarni}, {Kurk}, {Lutz}, {Nelson}, {Momcheva}, {Rosario},
  {Saglia}, {Seitz}, {Tacconi}, \& {van Dokkum}}]{Wisnioski15}
{Wisnioski}, E., {F{\"o}rster Schreiber}, N.~M., {Wuyts}, S., {et~al.} 2015,
  \apj, 799, 209

\bibitem[{{Wuyts} {et~al.}(2010){Wuyts}, {Cox}, {Hayward}, {Franx},
  {Hernquist}, {Hopkins}, {Jonsson}, \& {van Dokkum}}]{Wuyts10}
{Wuyts}, S., {Cox}, T.~J., {Hayward}, C.~C., {et~al.} 2010, \apj, 722, 1666

\bibitem[{{Zolotov} {et~al.}(2015){Zolotov}, {Dekel}, {Mandelker}, {Tweed},
  {Inoue}, {DeGraf}, {Ceverino}, {Primack}, {Barro}, \& {Faber}}]{Zolotov15}
{Zolotov}, A., {Dekel}, A., {Mandelker}, N., {et~al.} 2015, \mnras, 450, 2327

\end{thebibliography}

\end{document}